\documentclass[aps,prd,10pt,notitlepage,superscriptaddress,nofootinbib,numbers,showkeys]{revtex4-2}

\usepackage[utf8]{inputenc}
\usepackage[T1]{fontenc}
\usepackage{lmodern}
\usepackage{microtype}
\usepackage{amsmath,amssymb,amsfonts,mathrsfs}
\usepackage{bm}
\usepackage{graphicx}
\usepackage[dvipsnames]{xcolor}
\usepackage{hyperref}
\hypersetup{colorlinks=true,citecolor=Purple,linkcolor=Purple,urlcolor=Purple,breaklinks=true}
\usepackage[capitalize]{cleveref}
\usepackage{tikz}

\definecolor{lime}{HTML}{A6CE39}
\DeclareRobustCommand{\orcidicon}{%
	\begin{tikzpicture}
		\draw[lime, fill=lime] (0,0) 
		circle [radius=0.16] 
		node[white] {{\fontfamily{qag}\selectfont \tiny ID}};
		\draw[white, fill=white] (-0.0625,0.095) 
		circle [radius=0.007];
	\end{tikzpicture}
	\hspace{-2mm}
}

\foreach \x in {A, ..., Z}{%
	\expandafter\xdef\csname orcid\x\endcsname{\noexpand\href{https://orcid.org/\csname orcidauthor\x\endcsname}{\noexpand\orcidicon}}
}

\newcommand\orcidJonathan{{\href{https://orcid.org/0000-0001-9291-0893}{\orcidicon}}}
\newcommand\orcidEdson{{\href{https://orcid.org/0000-0001-9929-5977}{\orcidicon}}}
\newcommand\orcidcelio{{\href{https://orcid.org/0000-0002-1266-2218}{\orcidicon}}}
\newcommand\orcidmca{{\href{https://orcid.org/0009-0009-8703-6092}{\orcidicon}}}
\newcommand\orcidfurtado{{\href{https://orcid.org/0000-0002-1273-519X}{\orcidicon}}}

\begin{document}

\newcommand{\Del}{\nabla}

\title{Junction Conditions, Radial Stability, Thermodynamics, Optical Geometry and Appearance of Polymer-Quintessence Thin-Shell Wormholes}

\author{Jonathan A. Rebou\c{c}as\orcidJonathan\!\!}
\email{jalvesreboucas@ifce.edu.br}
\affiliation{Instituto Federal de Educa\c{c}\~ao, Ci\^encia e Tecnologia do Cear\'a (IFCE), Iguatu, Brazil}

\author{Edson Otoniel\orcidEdson\!\!}
\email{edson.otoniel@ufca.edu.br}
\affiliation{Universidade Federal do Cariri (UFCA), Instituto de Forma\c{c}\~ao de Educadores - IFE,  R. Oleg\'ario Emidio de Araujo S/N, Brejo Santo - CE, 63.260-000 - Brazil}

\author{M. C. Ara\'{u}jo\orcidmca}
\email{michelangelo.araujo@ufca.edu.br}
\affiliation{Universidade Federal do Cariri, Av. Tenente Raimundo Rocha, \\
Cidade Universit\'{a}ria, Juazeiro do Norte, Cear\'{a}, 63048-080, Brazil}

\author{J. Furtado\orcidfurtado}
\email{job.furtado@ufca.edu.br}
\affiliation{Universidade Federal do Cariri, Av. Tenente Raimundo Rocha, \\
Cidade Universit\'{a}ria, Juazeiro do Norte, Cear\'{a}, 63048-080, Brazil}

\author{Celio R. Muniz\orcidcelio}
\email{celio.muniz@uece.br}
\affiliation{Universidade Estadual do Cear\'a (UECE), Faculdade de Educa\c{c}\~ao, Ci\^encias e Letras de Iguatu, Av. D\'ario Rabelo s/n, Iguatu - CE, 63.500-00 - Brasil}

\begin{abstract}
Thin-shell wormholes built from effective black hole geometries are sensitive not only to the lapse function but also to the choice of areal radius. We construct a reflection-symmetric thin-shell wormhole from the positive-lapse sector of a polymer black hole surrounded by Kiselev-type quintessence and keep the nonareal angular function throughout the junction, stability, thermodynamic, and optical analyses. The Israel junction conditions give a negative surface energy density for every static throat on the positive branch, while the tangential null and intrinsic strong energy combinations are controlled by the local lapse slope. The radial dynamics is written as an effective-potential problem in which the nonareal sector produces a momentum-flux term and modifies the local stability criterion for surface equations of state with explicit radius dependence. For the sampled calibrated configurations, the linear barotropic and variable phantomlike closures remain locally unstable, whereas the variable Chaplygin gas admits finite linear radial stability windows. The same geometric correction also modifies the local first-law balance and shell entropy bookkeeping, while the optical analysis shows that cross-throat propagation generates additional inner image branches despite the wormhole and black hole geometries sharing the same exterior critical curve. These results identify how polymer corrections and a quintessence environment jointly reorganize the matter content, radial response, thermodynamic bookkeeping, and optical appearance of the resulting thin-shell wormhole.
\end{abstract}


\maketitle
\tableofcontents

\section{Introduction}\label{sec:introduction}
Compact strong-gravity geometries remain one of the most useful arenas for testing how classical general relativity, effective matter sectors, and quantum-inspired corrections constrain the admissible structure of spacetime. In this setting, wormhole configurations are valuable not only because they probe nontrivial topology, but also because they sharpen the relation between causal accessibility, effective stress tensors, and the observational distinction between horizonless compact objects and black holes. Recent work on ringing, lensing, shadows, photon-ring structure, and image reconstruction has reinforced this perspective by showing that optical and wave-dynamical diagnostics can encode features that are sensitive to the underlying throat geometry \cite{Konoplya2016WormholesVersus,Wang2020NovelShadows,Luo2023ObservationalAppearance,Tan2025AsymmetricThin}. For this reason, the wormhole problem continues to serve as a controlled theoretical laboratory for asking how far compact-object phenomenology can be extended beyond standard black hole exteriors without losing mathematical consistency.

Within this broader program, thin-shell wormholes provide an especially economical construction. Instead of specifying an exotic matter distribution throughout the bulk, one glues two spacetime regions across a timelike hypersurface and localizes the nontrivial source on the shell itself. The junction formalism of Israel gives the surface stress tensor in terms of the jump of the extrinsic curvature \cite{Israel1966}, while the effective-potential analysis of Poisson and Visser translates linearized radial perturbations into a local stability test \cite{PoissonVisser1995}. Since then, the same basic framework has been applied to plane-symmetric, cylindrical, rotating, higher-dimensional, regular, and modified-gravity backgrounds, among others \cite{Lemos2008PlaneSymmetric,Sharif2013MechanicalStability,Ovgun2016RotatingThin,Mehdizadeh2015HigherDimensional,Rahaman2009ThinShell,Forghani2018ThinShells,Liu2020TraversableThin,Tangphati2019ThinShell}. This body of work makes clear that the existence and dynamical viability of a thin-shell wormhole are controlled jointly by the seed geometry and by the constitutive response of the matter confined to the throat.

The surface equation of state is therefore not a secondary ingredient. Linear barotropic closures are often used to isolate the geometric part of the perturbation problem, but a large literature has shown that explicit radius dependence and nonlinear density dependence can alter the stability criterion substantially. Thin-shell wormholes supported by Chaplygin-type matter, generalized Chaplygin sectors, phantomlike models, and related nonlinear fluids have been investigated in several gravitational settings \cite{EiroaSimeone2007,Eiroa2009ThinShell,Bejarano2011DilatonThin,Eiroa2012ThinShell,Sharif2013SphericalThin,Varela2015,Kuhfittig2010StabilityThin,Azam2016BornInfeld}. In parallel, the search for shells that soften the energy-condition violation or modify its interpretation has motivated studies of thin-shells with unusual matter content and of cases where the background theory changes the junction balance itself \cite{Forghani2019ThinShell}. The lesson is that different equations of state do not merely decorate a fixed stability test: they reorganize the local perturbation problem and can expose qualitatively different equilibrium behavior even when the bulk geometry is held fixed.

The field has also moved well beyond a purely mechanical interpretation. Thermodynamic stability criteria, generalized first-law formulations, and connections between shell acceleration and effective temperature have all become relevant to the modern thin-shell literature \cite{Forghani2018ThermodynamicStability,Eiroa2024DynamicalThermodynamical,LoboRodrigues2026Thermo,LoboRodrigues2026Unruh}. At the same time, test-particle and photon dynamics on thin-shell backgrounds have provided complementary information about orbit classes, turning points, and observational appearance \cite{DiemerSmolarek2013,Wang2020NovelShadows,Luo2023ObservationalAppearance,Tan2025AsymmetricThin,Mazharimousavi2024GenericSpherically}. These developments are important for the present work because they show that a seed geometry useful for a thin-shell construction should be assessed not only by the sign of the shell density or by the local curvature of a radial potential, but also by the way it organizes temperature scales, horizon bookkeeping, and null-geodesic structure.

A natural class of seed metrics for this agenda comes from quantum-inspired or effective black hole geometries in which the classical Schwarzschild singularity is replaced or softened by polymer or loop-gravity corrections. Polymer black hole models have been studied as effective metrics, as examples of singularity resolution or bounce structure, and as laboratories for geodesics, thermodynamics, and phenomenological constraints \cite{Achour2018PolymerSchwarzschild,Livine2012EntropyClassical,Liu2021ExtendedGeometry,Liu2022SolarSystem,Mele2021QuantumCorrected,Tu2023PeriodicOrbits,Chen2024MotionSpinning,Mnch2022GenericFeatures}. This activity has made polymer geometries particularly attractive for thin-shell applications, because they preserve a relatively simple static and spherically symmetric form while introducing genuinely new scales into the horizon and orbit structure. A closely related recent line of work has already shown that entropy-induced effective black hole geometries can be used systematically as thin-shell seeds, with the stability problem then depending on the interplay between the deformed bulk metric and the shell equation of state \cite{Rebouas2026ThinShell3}.

An additional ingredient of direct relevance here is a surrounding quintessence-like environment. Kiselev's construction remains a standard phenomenological way of encoding such a contribution in a static lapse \cite{Kiselev2003}, and the resulting backgrounds have supported analyses of lensing, thermodynamic relations, geodesics, evaporation, and interpretive limits of the effective fluid description \cite{Younas2015StrongGravitational,Majeed2015ThermodynamicRelations,Pedraza2020GeodesicsHayward,Visser2019KiselevBlack,Wu2026KiselevBlack}. The direct source geometry for the present manuscript is the hybrid polymer-quintessence black hole introduced by Ara\'ujo \textit{et al.} \cite{Araujo2025}, who combined a polymer black hole sector with a Kiselev term and then studied the resulting thermodynamics, greybody factors, null geodesics, and shadows. That geometry is a particularly interesting seed for a thin-shell problem because the polymer correction does not merely deform the lapse: it also changes the angular sector, so that the radial coordinate is no longer the areal radius when the polymer scale is nonzero.

The closest direct predecessor to the present construction is the polymer thin-shell analysis of Javed \textit{et al.} \cite{Javed2022}, where a polymer black hole seed was combined with barotropic, variable phantomlike, and variable Chaplygin shell models. That study is methodologically important for the present manuscript because it shows how the same family of shell equations of state can lead to distinct stability outcomes even within a single quantum-inspired background. At the same time, recent work by the present authors on black hole-in-void thin-shells has highlighted how environmental structure can enrich the thermodynamic interpretation of the shell \cite{ReboucasOtonielLobo2026Voids}. Taken together, these nearby studies make the current question well posed: one would like to know what changes when the thin-shell program is transported from the pure polymer seed to the polymer-quintessence hybrid geometry, and when the angular sector itself forces a nonareal treatment of the shell.

This is precisely where the main gap lies. Although quintessence-supported thin-shell wormholes and related extensions have already been explored in other settings \cite{Banerjee2016StabilityDimensional}, and although polymer or regular backgrounds have repeatedly been used as thin-shell seeds \cite{Jusufi2016QuantumCorrected,Rahaman2009ThinShell,Alshal2019LinearizedStability,Alshal2024LinearizedStability}, to the best of our knowledge, a reflection-symmetric thin-shell construction based directly on the hybrid polymer-quintessence metric of Ref.~\cite{Araujo2025} has not been worked out with its full nonareal angular sector kept explicit. That point matters because once the areal radius is no longer identified with the seed radial coordinate, the surface conservation law, the shell internal-energy bookkeeping, and the null-geodesic potential do not reduce to their familiar areal-gauge forms. A naive transplantation of standard thin-shell formulas would therefore miss the momentum-flux term generated by the angular sector and would blur the distinction between the hybrid geometry and its Schwarzschild, Schwarzschild-Kiselev, or pure-polymer limits.

In this paper we construct a reflection-symmetric thin-shell wormhole by gluing two copies of a connected positive-lapse region of the polymer-quintessence geometry of Ref.~\cite{Araujo2025}. Using the Israel junction formalism, we derive the induced metric, orthonormal extrinsic curvatures, surface density, tangential pressure, and the corresponding energy-condition combinations. We then reduce the radial dynamics to an effective potential that retains both the nonareal flux contribution and the explicit radius dependence of the shell equation of state, and we specialize the resulting local stability criterion to a calibrated linear barotropic law, a variable phantomlike model, and a variable Chaplygin gas following the constitutive choices that connect most directly with the polymer thin-shell literature \cite{Javed2022,Varela2015,EiroaSimeone2007}. Because the same seed geometry also admits a meaningful optical and thermodynamic analysis, we further examine the local shell
temperature, the modified first-law balance, the null-geodesic potential, the corresponding embedding diagrams, and a simplified thin-disk radiative-transfer model that enables a direct, simplified optical comparison with the corresponding black hole geometry.

The physical significance of this program is twofold. First, it clarifies which features of the thin-shell problem follow from the hybrid seed geometry itself, and which follow from the material response of the shell. Second, it shows how polymer corrections and a quintessence-like environment can alter the standard thin-shell bookkeeping at the level of flux balance, horizon comparison, and light propagation without requiring a change of junction formalism. In this sense, the present construction is not simply another application of an established method to a new metric; it tests how a nonareal effective geometry reorganizes the relation between local throat dynamics, thermodynamic interpretation, and optical structure.

This paper is organized as follows. Section~\ref{sec:geometry} introduces the polymer-quintessence seed geometry and its relevant limits. In Sec.~\ref{sec:construction}, we perform the reflection-symmetric cut-and-paste construction and derive the induced throat geometry. Section~\ref{sec:junction} obtains the orthonormal extrinsic curvatures, surface stress tensor, and the energy-condition combinations. In Sec.~\ref{sec:dynamics}, we derive the surface conservation law, formulate the effective radial potential, and obtain the general local stability criterion, while Sec.~\ref{sec:eos} specializes the analysis to the three shell equations of state. Section~\ref{sec:thermodynamics} discusses the thermodynamic bookkeeping of the static shell. Section~\ref{sec:optics} develops the null-geodesic construction, the embedding diagram, the optical potential, and the corresponding thin-disk optical appearance images. Finally, Sec.~\ref{sec:conclusion} summarizes the main results and their limitations.

\section{Polymer-quintessence static geometry}\label{sec:geometry}

We begin this section by briefly reviewing the polymer black hole, which represents a quantum extension of the Schwarzschild spacetime within the framework of Loop Quantum Gravity (LQG) \cite{Bodendorfer:2019nvy,Bodendorfer:2019cyv,Brahma:2020eos,Tu2023PeriodicOrbits}. In general, the metric describing this black hole (BH) can be written as
\begin{eqnarray}\label{elemlinhalqggeral}
\mathrm{d}s^2 = -8A_{\lambda} M_{B}^{2} \mathcal{A}(r) \mathrm{d}t^2 + \frac{\mathrm{d}r^2}{8A_{\lambda} M_{B}^{2} \mathcal{A}(r)} + H(r) \left(d\theta^{2}+\sin^{2}\theta d\phi^{2}\right),
\end{eqnarray}
where
\begin{eqnarray}
\mathcal{A}(r) &=& \frac{1}{H(r)}\left(1 + \frac{r^2}{8A_{\lambda} M_{B}^{2}}\right) \left(1 - \frac{2M_B}{\sqrt{8A_{\lambda} M_{B}^{2} + r^2}}\right)
\end{eqnarray}
and
\begin{eqnarray}
H(r) &=& \frac{512A_{\lambda}^{3} M_{B}^{4} M_{W}^{2} + \left(r + \sqrt{8A_{\lambda} M_{B}^{2} + r^2}\right)^6}{8\sqrt{8A_{\lambda} M_{B}^{2} + r^2}\left(\sqrt{8A_{\lambda} M_{B}^{2} + r^2} + r\right)^3}.
\end{eqnarray}
Here, $A_{\lambda} \equiv (\lambda/(M_B M_W))^{2/3}/2$ is a dimensionless parameter defined in terms of two Dirac observables, $M_B$ and $M_W$, together with the quantum parameter $\lambda$, which has mass dimension $2$ and is associated with holonomy corrections in LQG \cite{Bodendorfer:2019nvy,Bodendorfer:2019cyv}. It is worth emphasizing that, in this effective quantum spacetime, the classical singularity of the Schwarzschild black hole is replaced by a quantum bounce connecting a black hole region to a white hole region, with $M_B$ and $M_W$ representing the masses of the black hole and white hole, respectively. This occurs when the function $H$ reaches its minimum at $r=0$ \cite{Bodendorfer:2019nvy,Bodendorfer:2019cyv,Brahma:2020eos}, closely resembling the quantum bounce scenario encountered in loop quantum cosmology \cite{Ashtekar:2011ni,Ashtekar:2006wn,Ashtekar:2009vc,Ashtekar:2015dja,Oriti:2016ueo,Diener:2017lde}.

In this work, our starting point is the static and spherically symmetric line element introduced in Ref.~\cite{Araujo2025}:
\begin{equation}\label{eq111111}
ds^{2}=-F(r)dt^{2}+\frac{dr^{2}}{F(r)}+H(r)\left(d\theta^{2}+\sin^{2}\theta d\phi^{2}\right),
\end{equation}
where
\begin{equation}\label{eq:lapse}
F(r)=\frac{4\ell^{2}-2M\sqrt{4\ell^{2}+r^{2}}+r^{2}}{\ell^{2}+r^{2}}-\frac{c_{q}}{r^{3w_{q}+1}},
\end{equation}
and
\begin{equation}\label{eq}
H(r)=r^{2}+\ell^{2},\qquad \ell=(\lambda M)^{1/3}.
\end{equation}
The polymer sector in Eqs.~\eqref{eq:lapse} and \eqref{eq} is obtained by restricting the effective metric in Eq.~\eqref{elemlinhalqggeral} to the symmetric-bounce sector $M_B=M_W=M$. Under this identification, $A_{\lambda}=(\lambda/M^{2})^{2/3}/2$ and $8A_{\lambda}M^{2}=4\ell^{2}$; substituting these relations into $\mathcal{A}(r)$ and $H(r)$ gives the functions displayed above without an additional coordinate transformation. Cosmological effects are then incorporated phenomenologically through the Kiselev radial dependence, characterized by the coupling parameter $c_q$ and the equation-of-state parameter $w_q$ \cite{Kiselev2003}. It is worth emphasizing that this effective model enabled the authors of Ref.~\cite{Araujo2025} to identify a broad range of parameter combinations $(\lambda,c_q,w_q)$ for which the predicted BH shadow sizes are consistent with the observational data for the black holes Sagittarius A* and M87*.

For suitable values of $(\lambda,c_q,w_q)$, the hybrid lapse can possess distinct positive roots delimiting a static patch, including an inner black-hole-type root and an outer cosmological-type root. Their number, ordering, and even their existence are parameter dependent and must therefore be determined for each chosen sector~\cite{Araujo2025}. The Kiselev contribution makes the hybrid geometry singular at the origin and destroys the asymptotic flatness of the pure-polymer solution, but these two effects should not be conflated. For $-1<w_q<-1/3$, its leading contributions to the Ricci and Kretschmann scalars scale respectively as $\mathcal{O}\!\left(r^{-3(1+w_q)}\right)$ and $\mathcal{O}\!\left(r^{-6(1+w_q)}\right)$ at large radius and therefore vanish as $r\to\infty$; the asymptotic obstruction lies in the metric behavior rather than in a curvature divergence.

Since we are working in geometrized units, we have $[M]=[\ell]=\mathcal{L}$, $[\lambda]=\mathcal{L}^{2}$, and $[c_q]=\mathcal{L}^{3w_q+1}$, where $\mathcal{L}$ denotes the dimension of length. Moreover, $0\leq\lambda\leq M^{2}$ in order to satisfy the existence condition for a pure polymer BH \cite{Tu2023PeriodicOrbits,Araujo2025}. We restrict the phenomenological Kiselev parameter to the customary quintessence interval $-1<w_q<-1/3$, while treating it here as a local lapse-shape parameter rather than as a cosmological-fluid fit; $c_q>0$, and $c_q=0$ recovers the pure polymer solution. Additionally, we impose the local geometric requirement $F(a_0)>0$ independently for each proposed static throat, where $a_0>0$ is an arbitrary length scale. It is also worth noting that the boundary value $w_q=-1$ produces a cosmological-constant-like lapse term, whereas $w_q=-1/3$ makes the Kiselev contribution a constant lapse shift. Although these boundary values are of considerable theoretical interest, they lie outside the open interval adopted in this work and will not be considered in the present analysis. Furthermore, the choice $\ell=0$ recovers the Schwarzschild-Kiselev solution, whereas imposing $c_q=0$ thereafter yields the Schwarzschild solution. By contrast, imposing only $c_q=0$ leaves the non-areal polymer angular sector unchanged and, as will be shown below, does not eliminate the associated junction-flux correction.

\begin{figure*}[htp!]
    \centering
    \includegraphics[width=0.98\linewidth]{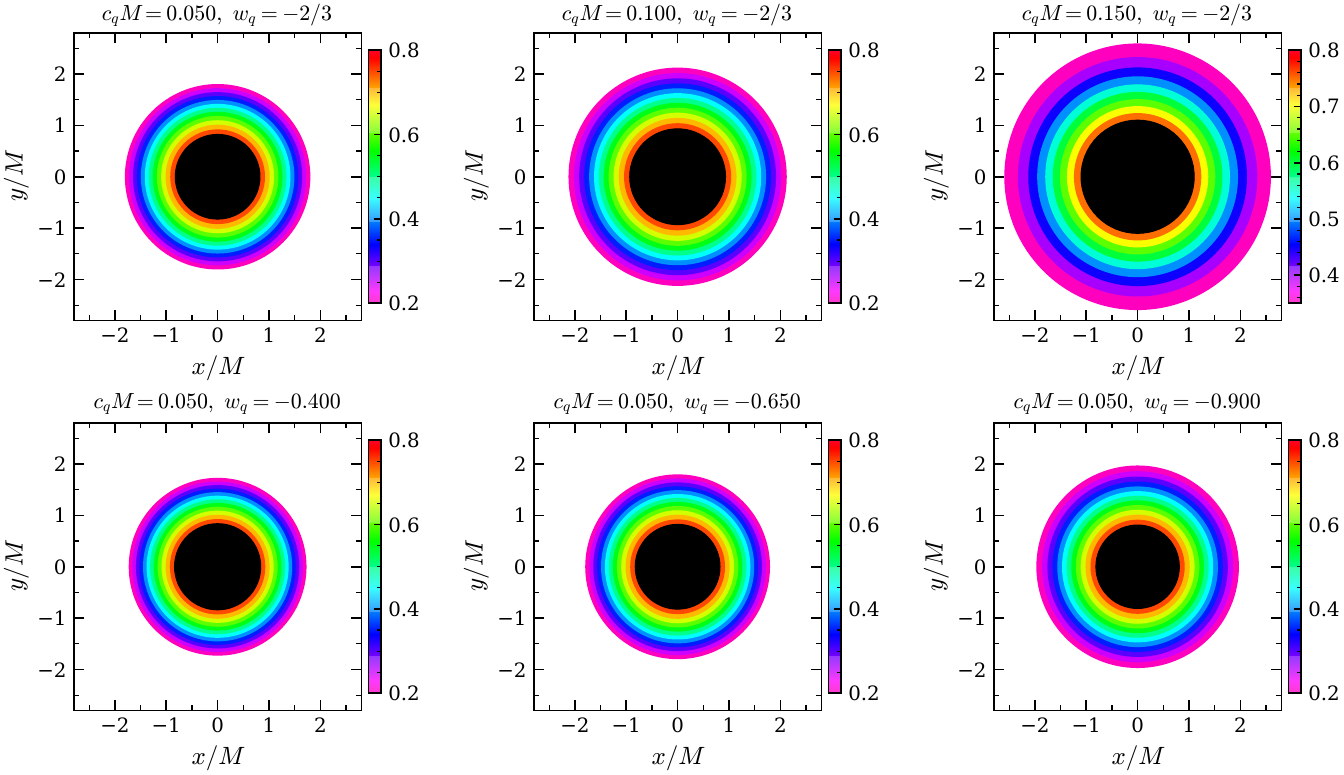}
    \caption{Selected positive-root parameter families for the polymer-quintessence seed geometry. Each colored disk is the smallest positive root of $g^{rr}=F(r)=0$ for one sampled value of $\lambda/M^2$, rather than a spatial shell or time evolution. In every panel $M=1$ and the color scale covers $0.2\leq\lambda/M^2\leq0.8$. In the first row $w_q=-2/3$ is fixed and the columns use the equally spaced values $c_qM=0.05,0.10,0.15$. In the second row $c_qM=0.05$ is fixed and the columns use the equally spaced values $w_q=-0.4,-0.65,-0.9$. Missing disks indicate sampled parameter values for which no positive selected root was found in the declared search range. The black disk marks the smallest selected radius in each plotted parameter family and is not an additional horizon.}
    \label{fig:horizon-family-2x3}
\end{figure*}

Figure~\ref{fig:horizon-family-2x3} indicates that the selected smallest positive root remains finite throughout the displayed families, but its radius is not controlled by the polymer parameter alone. In the first row, increasing $c_qM$ enlarges the whole annular family for fixed $w_q=-2/3$, so the Kiselev contribution shifts the inner static boundary outward for the sampled values. This behavior is consistent with Eq.~\eqref{eq:lapse}, where the quintessence term competes directly with the polymer-Schwarzschild part in the horizon condition $F(r)=0$.

The color ordering also shows that larger $\lambda/M^2$ corresponds to smaller selected radii inside each panel. This is the same qualitative direction as the pure-polymer inner-horizon dependence discussed in the source geometry, but here it should be read only as a selected-root statement inside the hybrid parameter sector because the Kiselev term can also introduce a distant cosmological root. The second row isolates the dependence on $w_q$ at fixed $c_qM=0.05$ and shows a milder variation of the selected-root family than the first-row variation in $c_qM$. Thus, within the displayed range, the Kiselev amplitude has the most visible effect on the selected-root scale, while $\lambda/M^2$ mainly orders the nested disks and $w_q$ changes the family more moderately.

\section{Symmetric throat inside the static patch}\label{sec:construction}

Let $\mathcal{M}_{+}$ and $\mathcal{M}_{-}$ be two identical copies of the portion of the seed spacetime lying in the same connected positive-lapse region and satisfying $r_{\pm}\geq a(\tau)$, with $a(\tau)>0$. Their timelike boundaries $\Sigma_{\pm}$ are the level sets $G_{\pm}(r_{\pm},\tau)=r_{\pm}-a(\tau)=0$. Identifying points of $\Sigma_{+}$ and $\Sigma_{-}$ with the same intrinsic coordinates produces the reflection-symmetric spacetime
\begin{equation}\label{eq:cut-and-paste-manifold}
 \mathcal{M}=\mathcal{M}_{+}\cup_{\Sigma}\mathcal{M}_{-},
 \qquad
 \Sigma\simeq\Sigma_{+}\simeq\Sigma_{-}.
\end{equation}
The identified hypersurface $\Sigma$ is the dynamical throat. Let its intrinsic coordinates be $\xi^i=(\tau,\theta,\phi)$, where $\tau$ is proper time measured by observers comoving with the shell. The functions $t_\pm(\tau)$ denote the static-time coordinates of $\Sigma$ as embedded in the two copies. The embedding on either side is
\begin{equation}\label{eq:embedding}
 X^{\mu}_{\pm}(\xi^i)=X^{\mu}_{\pm}(\tau,\theta,\phi)=\bigl(t_{\pm}(\tau),a(\tau),\theta,\phi\bigr).
\end{equation}
Here the signs label the two copies, and $X^\mu_{\pm}$ maps the intrinsic shell coordinates into the corresponding bulk coordinates. The pullback of the bulk metric is
\begin{equation}\label{eq:pullback}
 h_{ij}=g_{\mu\nu}\frac{\partial X^{\mu}}{\partial \xi^{i}}\frac{\partial X^{\nu}}{\partial \xi^{j}}.
\end{equation}
Proper-time normalization fixes the temporal component of the shell velocity through
\begin{equation}\label{eq:proper-time-normalization}
 -F(a)\dot t_\pm^{\,2}+\frac{\dot a^{2}}{F(a)}=-1,\qquad
 \dot t_\pm=\frac{\sqrt{F(a)+\dot a^2}}{F(a)}
\end{equation}
inside the chosen positive-lapse component. After that normalization, the induced line element is
\begin{equation}\label{eq:induced-metric}
 ds^{2}_{\Sigma}=-d\tau^{2}+H\bigl(a(\tau)\bigr)\left(d\theta^{2}+\sin^{2}\theta\,d\phi^{2}\right).
\end{equation}
Thus the physical throat radius is $R_\Sigma=\sqrt{H(a)}$, rather than $a$. On the branch $a>0$, $H'(a)=2a>0$, so increasing the coordinate radius also increases the throat area. The flare-out statement relevant to this surgery is the increase of area away from the minimum identified surface; it must not be inferred from $r$ as though $r$ itself were an areal coordinate.

With the outward orientation on each copy, the normalized-gradient prescription and its explicit evaluation give
\begin{equation}\label{eq:normal}
 n^{(\pm)}_{\mu}
 =\pm\left|g^{\alpha\beta}\partial_{\alpha}G_\pm\,\partial_{\beta}G_\pm\right|^{-1/2}\partial_{\mu}G_\pm
 =\pm\left(-\dot a,\frac{\sqrt{F(a)+\dot a^{2}}}{F(a)},0,0\right).
\end{equation}
The overdot denotes $d/d\tau$, and $n^{(\pm)}_\mu$ is a unit spacelike normal satisfying $n_\mu n^\mu=1$ and orthogonal to the shell velocity. The extrinsic curvature is
\begin{equation}\label{eq:extrinsic-definition}
 K^{(\pm)}_{ij}=-n^{(\pm)}_{\mu}\left(\frac{\partial^{2}X^{\mu}}{\partial\xi^{i}\partial\xi^{j}}+\Gamma^{\mu}_{\alpha\beta}\frac{\partial X^{\alpha}}{\partial\xi^{i}}\frac{\partial X^{\beta}}{\partial\xi^{j}}\right).
\end{equation}
The coefficients $\Gamma^{\mu}_{\alpha\beta}$ are those of the seed metric. In the shell's orthonormal frame, the independent covariant components are
\begin{equation}\label{eq:extrinsic-temporal}
 K^{(\pm)}_{\hat\tau\hat\tau}=\mp\frac{F'(a)+2\ddot a}{2\sqrt{F(a)+\dot a^{2}}}.
\end{equation}
The angular components in the same frame are
\begin{equation}\label{eq:extrinsic-angular}
 K^{(\pm)}_{\hat\theta\hat\theta}=K^{(\pm)}_{\hat\phi\hat\phi}
 =\pm\frac{H'(a)}{2H(a)}\sqrt{F(a)+\dot a^{2}}.
\end{equation}
Primes denote differentiation with respect to the displayed radial argument. The opposite signs between the two sides follow from the outward normals, while the relative temporal sign results from lowering the timelike orthonormal index.

\section{Orthonormal junction and surface matter}\label{sec:junction}

In the orthonormal basis $(\hat\tau,\hat\theta,\hat\phi)$, with induced metric $\eta_{\hat i\hat j}=\operatorname{diag}(-1,1,1)$, the covariant surface stress tensor of an isotropic tangential fluid is
\begin{equation}\label{eq:surface-tensor}
 S_{\hat i\hat j}=\operatorname{diag}(\sigma,p,p).
\end{equation}
The quantities $\sigma$ and $p$ are the surface energy density and tangential pressure. Writing $K=\eta^{\hat i\hat j}K_{\hat i\hat j}$ for the orthonormal trace and $[Q]=Q^{(+)}-Q^{(-)}$ for the jump of any junction quantity, the Israel--Lanczos equation in the same basis is \cite{Israel1966,PoissonVisser1995}
\begin{equation}\label{eq:lanczos}
 S_{\hat i\hat j}=-\frac{1}{8\pi}
 \left([K_{\hat i\hat j}]-\eta_{\hat i\hat j}[K]\right).
\end{equation}

Substitution of Eqs.~\eqref{eq:extrinsic-temporal} and \eqref{eq:extrinsic-angular} gives the dynamic surface density
\begin{equation}\label{eq:sigma-dynamic}
 \sigma=-\frac{H'(a)}{4\pi H(a)}\sqrt{F(a)+\dot a^{2}}.
\end{equation}
The corresponding tangential pressure is
\begin{equation}\label{eq:p-dynamic}
 p=\frac{1}{8\pi}\left[\frac{F'(a)+2\ddot a}{\sqrt{F(a)+\dot a^{2}}}+\frac{H'(a)}{H(a)}\sqrt{F(a)+\dot a^{2}}\right].
\end{equation}
For a static configuration at $a=a_0$, let $F_0=F(a_0)$ and $H_0=H(a_0)$. The timelike-throat requirement is $F_0>0$. The static density becomes
\begin{equation}\label{eq:sigma-static}
 \sigma_0=-\frac{a_0\sqrt{F_0}}{2\pi H_0}.
\end{equation}
Its pressure is
\begin{equation}\label{eq:p-static}
 p_0=\frac{1}{8\pi}\left[\frac{F'(a_0)}{\sqrt{F_0}}+\frac{2a_0\sqrt{F_0}}{H_0}\right].
\end{equation}
Because $a_0>0$, $H_0>0$, and $F_0>0$, every static throat on this branch has $\sigma_0<0$. The surface weak energy condition and dominant energy condition therefore fail without any further parameter test.
\begin{figure*}
    \centering
    \includegraphics[width=0.98\linewidth]{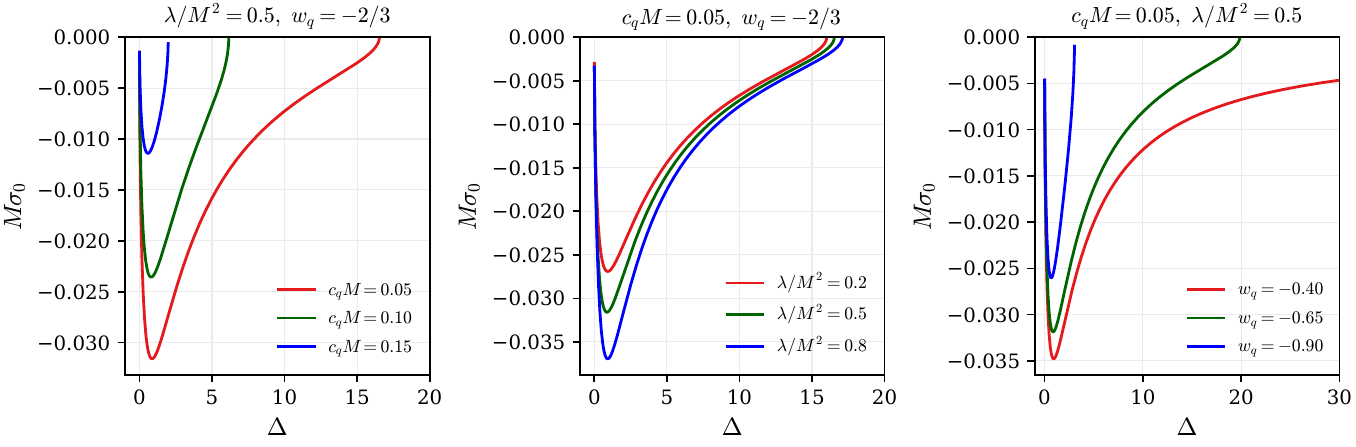}
    \caption{Static surface energy density $M\sigma_0$ as a function of the dimensionless throat offset $\Delta=(a_0-r_h)/M$. In the first panel, $\lambda/M^2=0.5$ and $w_q=-2/3$ are fixed, and the curves vary $c_qM=0.05,0.10,0.15$. In the second panel, $c_qM=0.05$ and $w_q=-2/3$ are fixed, and the curves vary $\lambda/M^2=0.2,0.5,0.8$. In the third panel, $c_qM=0.05$ and $\lambda/M^2=0.5$ are fixed, and the curves vary $w_q=-0.4,-0.65,-0.9$. In all panels $M=1$. The curves are restricted to the admissible positive-lapse region, and the offset labels distinct static throats outside the selected smallest positive root $r_h$.}
    \label{fig:shell-density}
\end{figure*}
The density profiles in Fig.~\ref{fig:shell-density} provide a direct numerical illustration of the sign statement following Eq.~\eqref{eq:sigma-static}. All displayed static throats have $M\sigma_0<0$, as expected from $a_0>0$, $H_0>0$, and $F_0>0$ on the retained branch. The curves approach the zero level only from below in the plotted windows, so the visual tendency toward small magnitude should not be interpreted as a removal of exotic matter at a finite throat.

The three panels separate changes in depth from changes in the admissible radial interval. Increasing $c_qM$ at fixed $\lambda/M^2=0.5$ and $w_q=-2/3$ reduces the depth of the negative minimum and shortens the visible positive-lapse branch. Varying $\lambda/M^2$ at fixed $c_qM=0.05$ produces a weaker separation of the curves, with larger $\lambda/M^2$ giving a more negative minimum on the displayed scale. In the $w_q$ sweep, the less negative value $w_q=-0.4$ extends much farther in $\Delta$, whereas the more negative branch terminates earlier. These trends are geometric rather than constitutive: before any equation of state is imposed, Eq.~\eqref{eq:sigma-static} ties the surface density to the local lapse and the nonareal area function.
The tangential null combination is
\begin{equation}\label{eq:nec-combination}
 \sigma_0+p_0=\frac{1}{8\pi\sqrt{F_0}}\left[F'(a_0)-\frac{2a_0F_0}{H_0}\right].
\end{equation}
For the intrinsic strong energy condition, the remaining relevant sum is
\begin{equation}\label{eq:sec-combination}
 \sigma_0+2p_0=\frac{F'(a_0)}{4\pi\sqrt{F_0}}.
\end{equation}
Consequently, neither the null nor the strong condition can be classified from the sign of $\sigma_0$ alone; both depend on the local lapse slope. Approaching a simple horizon from a static side drives $\sigma_0$ to zero from below, but generally makes $p_0$ diverge through the factor $F'(a_0)/\sqrt{F_0}$. A near-horizon throat is therefore not automatically a matter-free limit. At a degenerate root, the limiting ratio requires a separate expansion rather than substitution into these formulas.
We refer below to the null energy condition as NEC and to the intrinsic strong energy condition as SEC.

\begin{figure*}[htp!]
    \centering
    \includegraphics[width=0.98\linewidth]{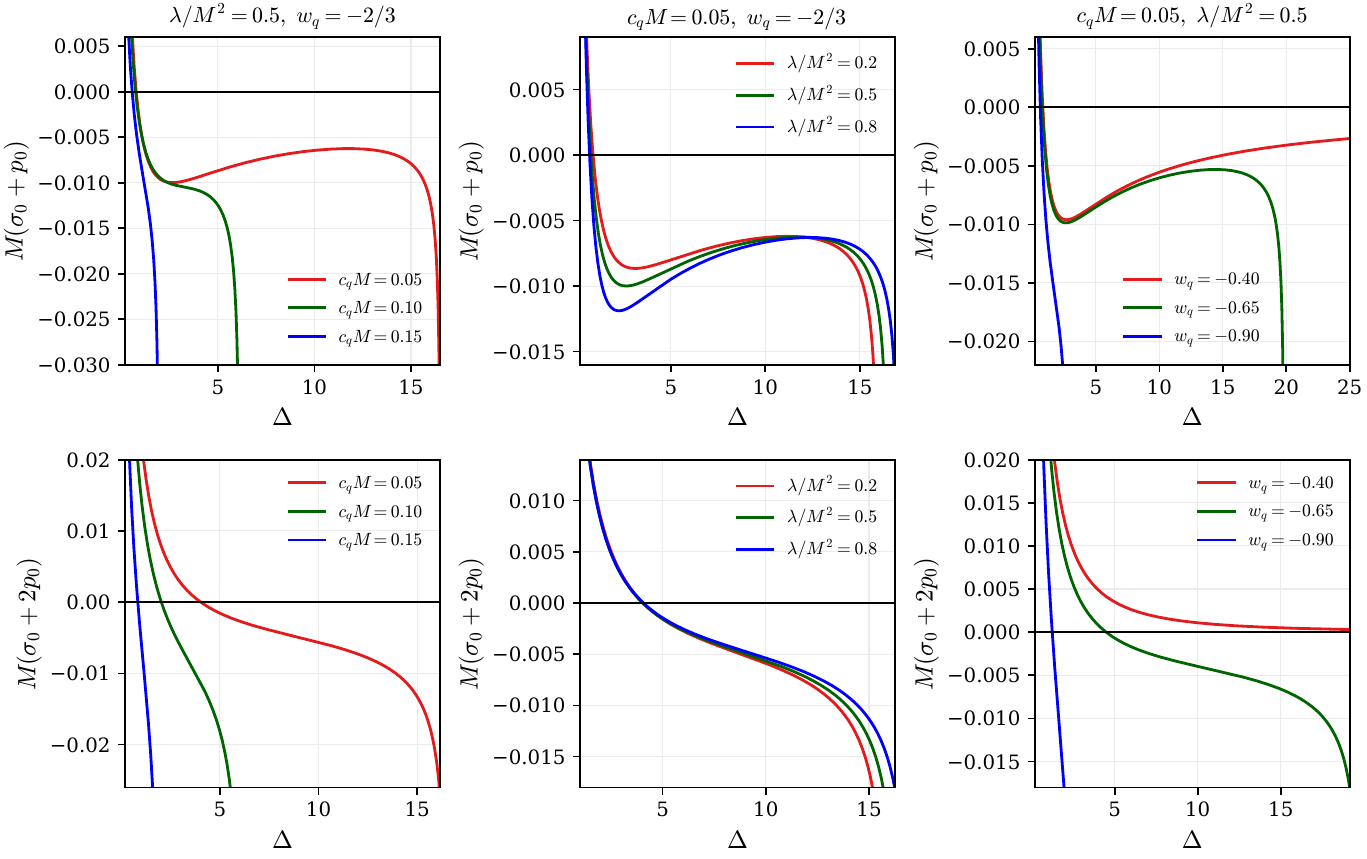}
    \caption{Static energy-condition combinations as functions of $\Delta=(a_0-r_h)/M$. The upper panels show the tangential NEC combination $M(\sigma_0+p_0)$, while the lower panels show the intrinsic SEC trace combination $M(\sigma_0+2p_0)$. In the first column, $\lambda/M^2=0.5$ and $w_q=-2/3$ are fixed, and the curves vary $c_qM=0.05,0.10,0.15$. In the second column, $c_qM=0.05$ and $w_q=-2/3$ are fixed, and the curves vary $\lambda/M^2=0.2,0.5,0.8$. In the third column, $c_qM=0.05$ and $\lambda/M^2=0.5$ are fixed, and the curves vary $w_q=-0.4,-0.65,-0.9$. In all panels $M=1$. Only admissible positive-lapse throats are plotted, and the horizontal line marks the zero level.}
    \label{fig:energy-conditions}
\end{figure*}

Figure~\ref{fig:energy-conditions} shows that the failure of the weak condition through $\sigma_0<0$ does not determine the sign of the tangential null or intrinsic strong combinations. In several panels both combinations are positive close to the selected horizon and then cross into negative values as the throat is moved outward. This is compatible with Eqs.~\eqref{eq:nec-combination} and \eqref{eq:sec-combination}, where the lapse slope competes with the density contribution and can dominate near a simple root of $F$.

The NEC row remains negative over most of the displayed intervals after the near-horizon transition. The $c_qM$ sweep shows that the branch with smaller $c_qM$ survives to larger $\Delta$, while larger $c_qM$ reaches the end of the admissible interval sooner. The $\lambda/M^2$ sweep produces nearly overlapping mid-range behavior, indicating weak sensitivity on the displayed scale except near the endpoint. In the $w_q$ sweep, the $w_q=-0.4$ curve has the widest radial extent, whereas the more negative choices terminate earlier.

The SEC trace row is more sensitive to the zero crossing. In the first column, increasing $c_qM$ moves the transition closer to the horizon and compresses the interval in which $M(\sigma_0+2p_0)$ is positive. In the second column, the three $\lambda/M^2$ curves remain close through the first crossing and separate mainly farther from the horizon. In the third column, the $w_q=-0.4$ branch approaches marginality from above within the displayed window, while the more negative branches cross and become negative. These are finite-window statements about the two plotted combinations; they do not imply simultaneous satisfaction of all shell energy conditions.

\section{Areal-radius flux and radial dynamics}\label{sec:dynamics}

The contracted junction identity gives a balance law for the shell. For $H=a^{2}+\ell^{2}$ it reduces to
\begin{equation}\label{eq:conservation}
 \frac{d\sigma}{da}=-\frac{2a}{H(a)}(\sigma+p)+\frac{\ell^{2}}{aH(a)}\sigma.
\end{equation}
The first term is the familiar dilution or work term, written with the true area $4\pi H$. The second is the momentum-flux contribution induced by the nonareal radial geometry. It survives in the pure-polymer limit $c_q\to0$ and vanishes only when $\ell\to0$, where $H\to a^{2}$ and the standard transparent-shell equation is recovered. Thus reciprocal temporal and radial metric coefficients do not, by themselves, justify dropping the flux term. The present equation also presumes $a>0$, the branch used throughout the construction.

Solving Eq.~\eqref{eq:sigma-dynamic} for the radial velocity yields an energy-form equation
\begin{equation}\label{eq:potential-equation}
 \dot a^{2}+V(a)=0.
\end{equation}
The effective potential appearing here is
\begin{equation}\label{eq:potential}
 V(a)=F(a)-4\pi^{2}\frac{H(a)^{2}}{a^{2}}\sigma(a)^{2}.
\end{equation}
For a general surface equation of state $p=p(\sigma,a)$, introduce the two local response coefficients
\begin{equation}\label{eq:eos-derivatives}
 \zeta=\left(\frac{\partial p}{\partial\sigma}\right)_{a},\qquad p_{,a}=\left(\frac{\partial p}{\partial a}\right)_{\sigma}.
\end{equation}
The coefficient $\zeta$ is a constitutive slope. On an exotic, effectively two-dimensional shell it is not automatically a causal bulk sound speed, while $p_{,a}$ records explicit radius dependence at fixed density. Differentiating the potential and using the full balance law gives
\begin{align}\label{eq:general-vpp}
 V''={}&F''-8\pi^{2}(\sigma+2p)^{2}
 -16\pi^{2}\sigma(1+2\zeta)(\sigma+p)\notag\\
 &+8\pi^{2}\frac{\ell^{2}}{a^{2}}\sigma^{2}(1+2\zeta)
 +16\pi^{2}\frac{H(a)}{a}\sigma p_{,a}.
\end{align}
All quantities in Eq.~\eqref{eq:general-vpp} are evaluated along the shell trajectory. At a calibrated static solution the junction equations ensure $V(a_0)=0$ and $V'(a_0)=0$. To state the linear test without confusing the perturbation with a physical parameter, write the radius multiplicatively as
\begin{equation}\label{eq:perturbation}
 a(\tau)=a_0\left[1+\varepsilon y(\tau)\right],\qquad |\varepsilon|\ll1.
\end{equation}
The constant $\varepsilon$ is a dimensionless bookkeeping amplitude, and $y$ is the dimensionless radial mode. This local amplitude should not be confused with the displacement parameter $\mathcal{E}$ used below in the stability plots, where $a_0=r_h+|\mathcal{E}|M$ labels a family of distinct calibrated static throats. For each value of $\mathcal{E}$, the perturbation parameter $\varepsilon$ remains infinitesimal and the stability test is the sign of $V''(a_0)$. The linearized radial equation is
\begin{equation}\label{eq:linear-mode}
 \ddot y+\frac{1}{2}V''(a_0)y=0.
\end{equation}
Hence $V''(a_0)>0$ is the local oscillatory-stability criterion, $V''(a_0)<0$ gives a growing mode, and the zero case is inconclusive at quadratic order. This is a local statement at a specified static throat; it neither establishes a globally traversable extension nor supplies stable parameter intervals without an independent scan.

\section{Thermodynamic bookkeeping of the static shell}\label{sec:thermodynamics}

The static construction also admits a local thermodynamic interpretation, but
the nonareal polymer sector changes the fixed-parameter first-law balance.  The
area of the throat is
\begin{equation}\label{eq:shell-area-thermo}
 A_\Sigma=4\pi H(a_0).
\end{equation}
This is the geometric area of the two-sphere carried by the shell.  It is not
$4\pi a_0^2$ unless the polymer scale is removed, so the thermodynamic
bookkeeping must use the same angular function that entered the junction
calculation.

For any nondegenerate Killing horizon $r_\alpha$ of the seed geometry, defined
by $F(r_\alpha)=0$, the usual static-patch temperature scale is
\begin{equation}\label{eq:horizon-temperature-thermo}
 T_\alpha=\frac{|F'(r_\alpha)|}{4\pi}.
\end{equation}
If the effective metric is assigned the standard area entropy, the associated
horizon entropy is $S_\alpha=\pi H(r_\alpha)$.  This area-law statement is a
geometric convention for the effective background; possible microscopic
polymer corrections to the entropy are not fixed by the present thin-shell
calculation.  In contrast with models whose static patch is fixed in advance
by two horizon boundaries, the present polymer--quintessence geometry can have
one or two positive nondegenerate roots, or lose one of them, depending on the
chosen values of $(M,\lambda,c_q,w_q)$.  The label $\alpha$ is therefore kept
generic: when a black-hole-like and a cosmological-like root coexist, the same
formula applies to each boundary separately, while in one-horizon sectors there
is no second horizon entropy to compare with the shell.  This is why the
thermodynamic bookkeeping below is written for an arbitrary admissible horizon
rather than specialized to a two-boundary phase space.

A static observer comoving with the throat has proper
acceleration magnitude $|F'(a_0)|/(2\sqrt{F_0})$.  Following the local
acceleration-temperature interpretation of thin shells, the natural shell
temperature is therefore
\begin{equation}\label{eq:shell-temperature-thermo}
 T_\Sigma=\frac{|F'(a_0)|}{4\pi\sqrt{F_0}}.
\end{equation}
This temperature is local and kinematical.  It does not, by itself, assert a
Planckian radiation spectrum; the particle-creation temperature requires a
separate ray-tracing condition, such as the exponential peeling behavior used
in the generalized Unruh analysis of dynamical thin-shell wormholes
\cite{LoboRodrigues2026Unruh}.

The internal energy of the surface fluid is defined as
$E_\Sigma=\sigma_0 A_\Sigma$.  Using Eq.~\eqref{eq:sigma-static}, this gives
\begin{equation}\label{eq:shell-internal-energy-thermo}
 E_\Sigma=-2a_0\sqrt{F_0}.
\end{equation}
The sign reflects the negative surface density of the symmetric junction, not
an independent assumption about the constitutive model.  The first-law form for
the matter on the throat follows the thin-shell thermodynamic framework of
Ref.~\cite{LoboRodrigues2026Thermo}:
\begin{equation}\label{eq:shell-first-law-thermo}
 T_\Sigma\,dS_\Sigma=dE_\Sigma+p_0\,dA_\Sigma.
\end{equation}
For a sequence of static throats at fixed seed parameters, Eqs.~\eqref{eq:p-static}
and \eqref{eq:shell-area-thermo} give
\begin{equation}\label{eq:fixed-parameter-first-law-thermo}
 \frac{dE_\Sigma}{da_0}+p_0\frac{dA_\Sigma}{da_0}
 =-\frac{2\ell^2\sqrt{F_0}}{H_0}.
\end{equation}
The right-hand side vanishes only in the areal-radius limit $\ell\to0$.  Thus,
unlike the transparent areal case, the shell entropy is not generally constant
along a fixed-parameter radial sequence when the polymer angular sector is
retained.  The same term is the thermodynamic counterpart of the flux
contribution already present in the surface balance law,
Eq.~\eqref{eq:conservation}; it measures the failure of the coordinate radius
to dilute the shell energy as an areal radius would.

Combining Eqs.~\eqref{eq:shell-temperature-thermo} and
\eqref{eq:fixed-parameter-first-law-thermo} yields
\begin{equation}\label{eq:fixed-parameter-entropy-gradient-thermo}
 \frac{dS_\Sigma}{da_0}
 =-\frac{8\pi\ell^2F_0}{H_0|F'(a_0)|}.
\end{equation}
This relation should be read locally inside a connected positive-lapse region
with $F_0>0$ and $F'(a_0)\neq0$.  It becomes inconclusive at an extremum of the
lapse, where the acceleration temperature vanishes, and it requires a separate
near-root expansion at degenerate horizons.  For a simple horizon approached
from the static side, the factor $F_0$ drives the fixed-parameter entropy
gradient to zero, while the pressure in Eq.~\eqref{eq:p-static} may still
diverge through $F'(a_0)/\sqrt{F_0}$.  A near-horizon shell is therefore not a
regular zero-work thermodynamic limit.

One may also vary a mass parameter along a family of static equilibria.  Let
$\upsilon$ denote the chosen external parameter and hold the remaining seed
parameters fixed according to the selected prescription.  Then
\begin{equation}\label{eq:parameter-first-law-thermo}
 T_\Sigma\,dS_\Sigma
 =-\frac{2\ell^2\sqrt{F_0}}{H_0}\,da_0
 +\left[-\frac{a_0F_{,\upsilon}(a_0)}{\sqrt{F_0}}
 +\frac{H_{,\upsilon}(a_0)F'(a_0)}{2\sqrt{F_0}}
 +\frac{a_0H_{,\upsilon}(a_0)\sqrt{F_0}}{H_0}\right]d\upsilon .
\end{equation}
Here $F_{,\upsilon}$ and $H_{,\upsilon}$ are partial derivatives at fixed coordinate
radius.  This compact notation is intentional: for the polymer--quintessence
metric the answer depends on whether one keeps $\lambda$, $\lambda/M^2$, or
another parameter combination fixed while varying $M$.  The thermodynamic
phase space is therefore not specified by the static junction equations alone.

The horizon entropies provide useful bookkeeping once such a parameter
prescription is chosen.  Differentiating the horizon condition gives
\begin{equation}\label{eq:horizon-shift-thermo}
 F'(r_\alpha)\,dr_\alpha+F_{,\upsilon}(r_\alpha)\,d\upsilon=0.
\end{equation}
Consequently, under the area-law convention,
\begin{equation}\label{eq:horizon-entropy-shift-thermo}
 dS_\alpha=\pi\left[H_{,\upsilon}(r_\alpha)
 -\frac{H'(r_\alpha)F_{,\upsilon}(r_\alpha)}{F'(r_\alpha)}\right]d\upsilon .
\end{equation}
Equations~\eqref{eq:parameter-first-law-thermo} and
\eqref{eq:horizon-entropy-shift-thermo} show how the shell entropy can be
compared with whichever horizon boundaries exist in the selected parameter
sector.  They do not constitute a Smarr relation, because the latter would
require a well-defined equilibrium curve $a_0(\upsilon)$, a choice of thermodynamic
variables, and a validated horizon first law for the effective
polymer--quintessence background.  The conclusion is therefore more limited
than in areal-radius void models \cite{ReboucasOtonielLobo2026Voids}: the
present shell has a well-defined local temperature and a traceable first-law
balance, but its entropy is controlled by the polymer flux term as well as by
any chosen horizon-parameter variation.

\section{Constitutive closures}\label{sec:eos}

We now specialize the general criterion to three surface equations of state used in the thin-shell literature. In every model, the equation is imposed on the dynamic shell before its constant is calibrated at a chosen static radius. The calibration constant is then held fixed while the shell is perturbed; recalibrating it along the motion would change the material model and invalidate the derivative test. Surface densities and pressures have dimension $\mathcal{L}^{-1}$ in geometrized units.

\subsection{Linear barotropic surface fluid}\label{sec:barotropic}

The dynamic linear law is
\begin{equation}\label{eq:eos-barotropic}
 p=w_s\sigma.
\end{equation}
The constant $w_s$ is dimensionless. In the source thin-shell application this closure is presented algebraically, without a universal numerical interval that must be inherited by a different seed geometry \cite{Javed2022}; the mathematical domain is therefore $w_s\in\mathbb{R}$. For the present surface model we likewise impose no bulk-fluid prior such as $0\leq w_s\leq1$, because the shell has negative energy density and its constitutive derivative cannot be identified uncritically with an ordinary three-dimensional sound speed. The adopted domain is instead the subset of real values generated by static calibration at admissible radii. No sampled interval is claimed in this analytic draft. The limit $w_s=0$ is a pressureless algebraic shell, whereas $w_s=-1$ saturates $\sigma+p=0$; neither limit guarantees compatibility with an arbitrary geometry. The law has no singular coefficient, but a static calibration is undefined if $\sigma$ vanishes, which excludes direct calibration at a horizon. Its appeal is diagnostic: it isolates the response to density while suppressing explicit radius dependence, as in the classic potential treatment of thin shells \cite{PoissonVisser1995}.

For this model the local response coefficients are
\begin{equation}\label{eq:barotropic-derivatives}
 \zeta=w_s,\qquad p_{,a}=0.
\end{equation}
Static compatibility fixes the otherwise free material constant to
\begin{equation}\label{eq:barotropic-calibration}
 w_s=\frac{p_0}{\sigma_0}.
\end{equation}
After this calibration, Eq.~\eqref{eq:general-vpp} becomes
\begin{align}\label{eq:barotropic-vpp}
 V''_{\mathrm B}(a_0)={}&F''(a_0)-8\pi^{2}(\sigma_0+2p_0)^{2}
 -16\pi^{2}(\sigma_0+2p_0)(\sigma_0+p_0)\notag\\
 &+8\pi^{2}\frac{\ell^{2}}{a_0^{2}}\sigma_0(\sigma_0+2p_0).
\end{align}
The last term is absent in areal gauge. Indeed, taking $\ell\to0$ after calibration reduces this expression to $F''+F'/a-F'^{,2}/F$, evaluated at the throat. That local limit is the standard transparent-shell barotropic result; setting $c_q\to0$ alone does not reach it because the polymer angular sector remains nonareal.

\begin{figure*}[htp!]
    \centering
    \includegraphics[width=0.98\linewidth]{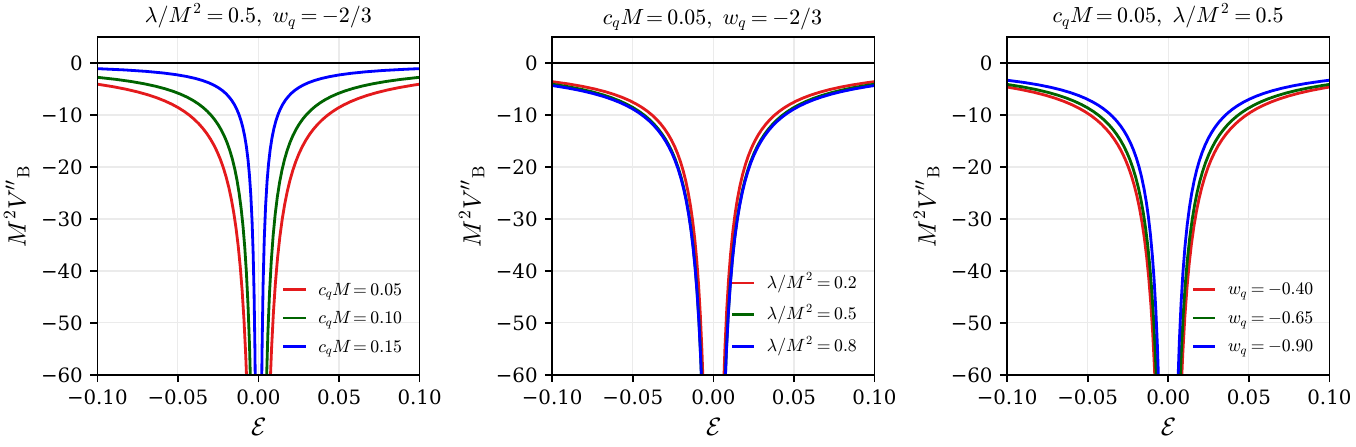}
    \caption{Linear barotropic stability function $M^2V''_{\mathrm{B}}(a_0)$ as a function of the dimensionless throat displacement $\mathcal{E}$, with $a_0=r_h+|\mathcal{E}|M$. In the first panel, $\lambda/M^2=0.5$ and $w_q=-2/3$ are fixed, and the curves vary $c_qM=0.05,0.10,0.15$. In the second panel, $c_qM=0.05$ and $w_q=-2/3$ are fixed, and the curves vary $\lambda/M^2=0.2,0.5,0.8$. In the third panel, $c_qM=0.05$ and $\lambda/M^2=0.5$ are fixed, and the curves vary $w_q=-0.4,-0.65,-0.9$. In all panels $M=1$, with $w_s$ calibrated at each static throat before the local perturbation test. The horizontal zero level separates stable configurations, $V''>0$, from unstable ones, $V''<0$.}
    \label{fig:stability-barotropic}
\end{figure*}

The barotropic scan in Fig.~\ref{fig:stability-barotropic} stays below the zero line for all displayed values of $\mathcal{E}$. According to the convention established by Eq.~\eqref{eq:linear-mode}, this means that the sampled calibrated barotropic throats are linearly unstable at quadratic order. The sharp downward behavior near $\mathcal{E}=0$ is a near-horizon feature of the local curvature test; it should not be read as the time evolution of one shell, because each point in the plot corresponds to a distinct static throat with a separately calibrated $w_s$.

The parameter sweeps mainly change the depth and width of the negative well. In the first panel, larger $c_qM$ raises the curves toward zero away from the near-horizon region, whereas smaller $c_qM$ gives a more negative response over a wider part of the plotted interval. The middle panel shows only weak separation among the $\lambda/M^2$ curves on this scale, so the polymer parameter does not create a positive stability window for the displayed barotropic family. In the third panel, the more negative value $w_q=-0.9$ lies closer to zero away from the center, but it still remains negative. Thus the scan is consistent with the analytic structure of Eq.~\eqref{eq:barotropic-vpp}: the calibrated linear density response and the nonareal correction do not overcome the destabilizing terms for these selected configurations.

\subsection{Variable phantomlike surface fluid}\label{sec:phantom}

The radius-dependent phantomlike closure is
\begin{equation}\label{eq:eos-phantom}
 p=A_P a^{-n}\sigma.
\end{equation}
The coefficient $A_P$ has physical dimension $\mathcal{L}^n$, while the exponent $n$ is dimensionless; the limit $n=0$ removes the explicit radial dependence.
Varela introduced this variable form with constant coefficients and showed that $n=0$ recovers a constant phantomlike law \cite{Varela2015}; it was subsequently used in the supplied polymer thin-shell source \cite{Javed2022}. The original analyses treat $n$ as real, although positive $n$ is singled out when one wants a coefficient that decays outward and in particular Schwarzschild stability sectors. That stability observation is not a fundamental domain and cannot be transported to the hybrid lapse. Mathematically, $n\in\mathbb{R}$, $A_P\in\mathbb{R}$, and $a>0$, with $[A_P]=\mathcal{L}^n$. The surface-adopted domain keeps all real $n$ for local analysis; the physically suggestive decaying-profile subcase is $n>0$. No numerical subset is sampled here. The sign of $A_P a^{-n}$ decides whether pressure and negative density have equal or opposite signs, so the word ``phantomlike'' labels the functional form rather than proving a particular energy-condition violation. The coefficient is regular for every finite positive throat. At $n=0$ the law coincides exactly with the barotropic closure, while $A_P=0$ is pressureless and generally cannot support a static shell whose junction pressure is nonzero.

The two response coefficients now read
\begin{equation}\label{eq:phantom-derivatives}
 \zeta=A_Pa^{-n},\qquad p_{,a}=-\frac{np}{a}.
\end{equation}
Compatibility with the chosen equilibrium fixes
\begin{equation}\label{eq:phantom-calibration}
 A_P=a_0^{n}\frac{p_0}{\sigma_0}.
\end{equation}
The corresponding stability curvature is
\begin{align}\label{eq:phantom-vpp}
 V''_{\mathrm P}(a_0)={}&F''(a_0)-8\pi^{2}(\sigma_0+2p_0)^{2}
 -16\pi^{2}(\sigma_0+2p_0)(\sigma_0+p_0)\notag\\
 &+8\pi^{2}\frac{\ell^{2}}{a_0^{2}}\sigma_0(\sigma_0+2p_0)
 -16\pi^{2}n\frac{H_0}{a_0^{2}}\sigma_0p_0.
\end{align}
The final term comes solely from explicit radial dependence and must not be absorbed into the density derivative. For $n=0$ both the calibration and the curvature reduce to the barotropic case. In areal gauge the criterion becomes $F''+F'/a-F'^{,2}/F+n(aF'+2F)/a^{2}$, showing separately which correction is constitutive and which is polymer-geometric.

\begin{figure*}[htp!]
    \centering
    \includegraphics[width=0.98\linewidth]{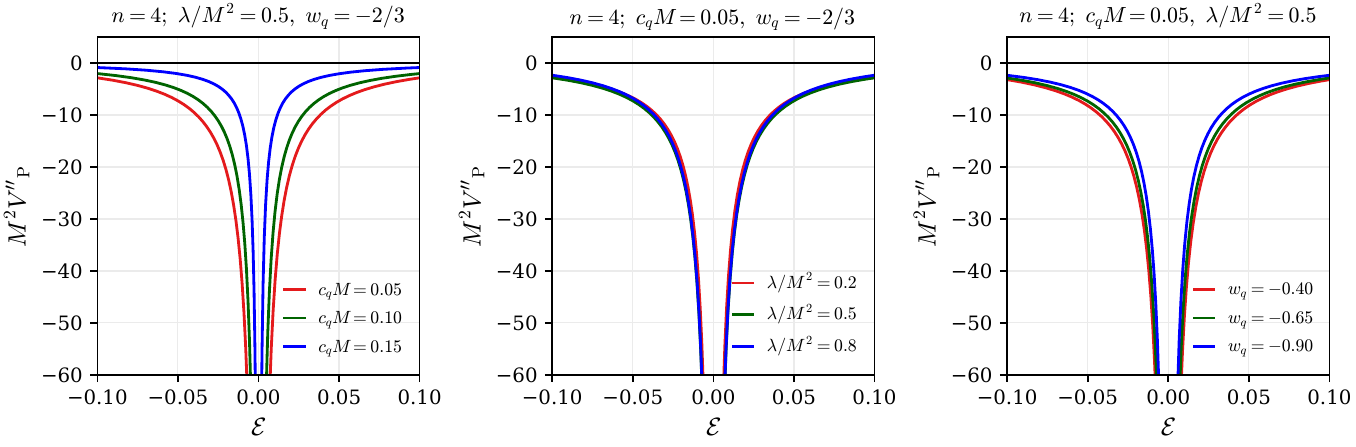}
    \caption{Variable phantomlike stability function $M^2V''_{\mathrm{P}}(a_0)$ as a function of the dimensionless throat displacement $\mathcal{E}$, with $a_0=r_h+|\mathcal{E}|M$. All panels use $n=4$. In the first column, $\lambda/M^2=0.5$ and $w_q=-2/3$ are fixed, and the curves vary $c_qM=0.05,0.10,0.15$. In the second column, $c_qM=0.05$ and $w_q=-2/3$ are fixed, and the curves vary $\lambda/M^2=0.2,0.5,0.8$. In the third column, $c_qM=0.05$ and $\lambda/M^2=0.5$ are fixed, and the curves vary $w_q=-0.4,-0.65,-0.9$. In all panels $M=1$, with $A_P$ calibrated at each static throat before the local perturbation test. The horizontal zero level separates stable configurations, $V''>0$, from unstable ones, $V''<0$.}
    \label{fig:stability-phantom}
\end{figure*}

Figure~\ref{fig:stability-phantom} shows that the variable phantomlike closure with the displayed value $n=4$ does not produce a positive $M^2V''_{\mathrm P}(a_0)$ branch in the sampled neighborhood of the selected smallest positive root. The curves have the same qualitative sign structure as the barotropic case in Fig.~\ref{fig:stability-barotropic}: they plunge downward near $\mathcal{E}=0$ and approach less negative values as $|\mathcal{E}|$ increases, but they do not cross the zero reference line within the plotted interval.

The comparison is useful because Eq.~\eqref{eq:phantom-vpp} differs from Eq.~\eqref{eq:barotropic-vpp} by the explicit radius-dependent term proportional to $n\sigma_0p_0$. For the present calibration and geometry, that additional contribution changes the scale only mildly and does not reverse the sign. The $c_qM$ and $w_q$ sweeps again show visible ordering away from the near-horizon region, while the $\lambda/M^2$ sweep remains comparatively compressed. Within the finite scan shown here, the variable phantomlike response therefore remains locally unstable even though the equation of state contains explicit radial dependence.

\subsection{Variable Chaplygin surface gas}\label{sec:chaplygin}

The third dynamic closure is
\begin{equation}\label{eq:eos-chaplygin}
 p=B_C\frac{a^{-n}}{\sigma}.
\end{equation}
The coefficient $B_C$ has physical dimension $\mathcal{L}^{n-2}$, and $n$ is dimensionless.
Its radius-dependent form follows Varela, while $n=0$ returns the ordinary Chaplygin relation used in earlier thin-shell analyses \cite{EiroaSimeone2007,Varela2015,Javed2022}. The mathematical domain is $n\in\mathbb{R}$, $B_C\in\mathbb{R}$, $a>0$, and $\sigma\neq0$, with $[B_C]=\mathcal{L}^{n-2}$. The nonzero-density restriction is essential rather than cosmetic: the pressure has a pole at $\sigma=0$. Our static branch has $\sigma<0$ in the interior of a positive-lapse patch, so it is locally admissible, but calibration directly at a horizon is forbidden and a horizon limit can make the pressure singular. In the conventional ordinary law $p=-A/\sigma$ with $A>0$ \cite{EiroaSimeone2007}; the present notation reproduces it through $B_C=-A<0$ when $n=0$. For the surface-adopted domain, however, the sign of $B_C$ is fixed by junction calibration rather than assumed beforehand. Positive $n$ again describes an outward-decaying explicit factor, not a mandatory physical range. No numerical subset is sampled. The formal value $B_C=0$ gives zero pressure away from the pole but cannot support a generic static junction.

For the variable Chaplygin law, differentiation at fixed radius and at fixed density gives
\begin{equation}\label{eq:chaplygin-derivatives}
 \zeta=-\frac{p}{\sigma},\qquad p_{,a}=-\frac{np}{a}.
\end{equation}
The static shell determines the constant through
\begin{equation}\label{eq:chaplygin-calibration}
 B_C=a_0^{n}\sigma_0p_0.
\end{equation}
The local potential curvature is therefore
\begin{align}\label{eq:chaplygin-vpp}
 V''_{\mathrm C}(a_0)={}&F''(a_0)-8\pi^{2}(\sigma_0+2p_0)^{2}
 -16\pi^{2}(\sigma_0-2p_0)(\sigma_0+p_0)\notag\\
 &+8\pi^{2}\frac{\ell^{2}}{a_0^{2}}\sigma_0(\sigma_0-2p_0)
 -16\pi^{2}n\frac{H_0}{a_0^{2}}\sigma_0p_0.
\end{align}
At $n=0$ this is the radius-independent Chaplygin criterion. Taking $\ell\to0$ yields $F''+(n+1)F'/a+(2n-4)F/a^{2}$ at the equilibrium radius. A further $c_q\to0$ recovers the Schwarzschild expressions of the original variable-EOS analysis, whereas $c_q\to0$ at finite $\ell$ deliberately retains both nonareal corrections.

\begin{figure*}[htp!]
    \centering
    \includegraphics[width=0.98\linewidth]{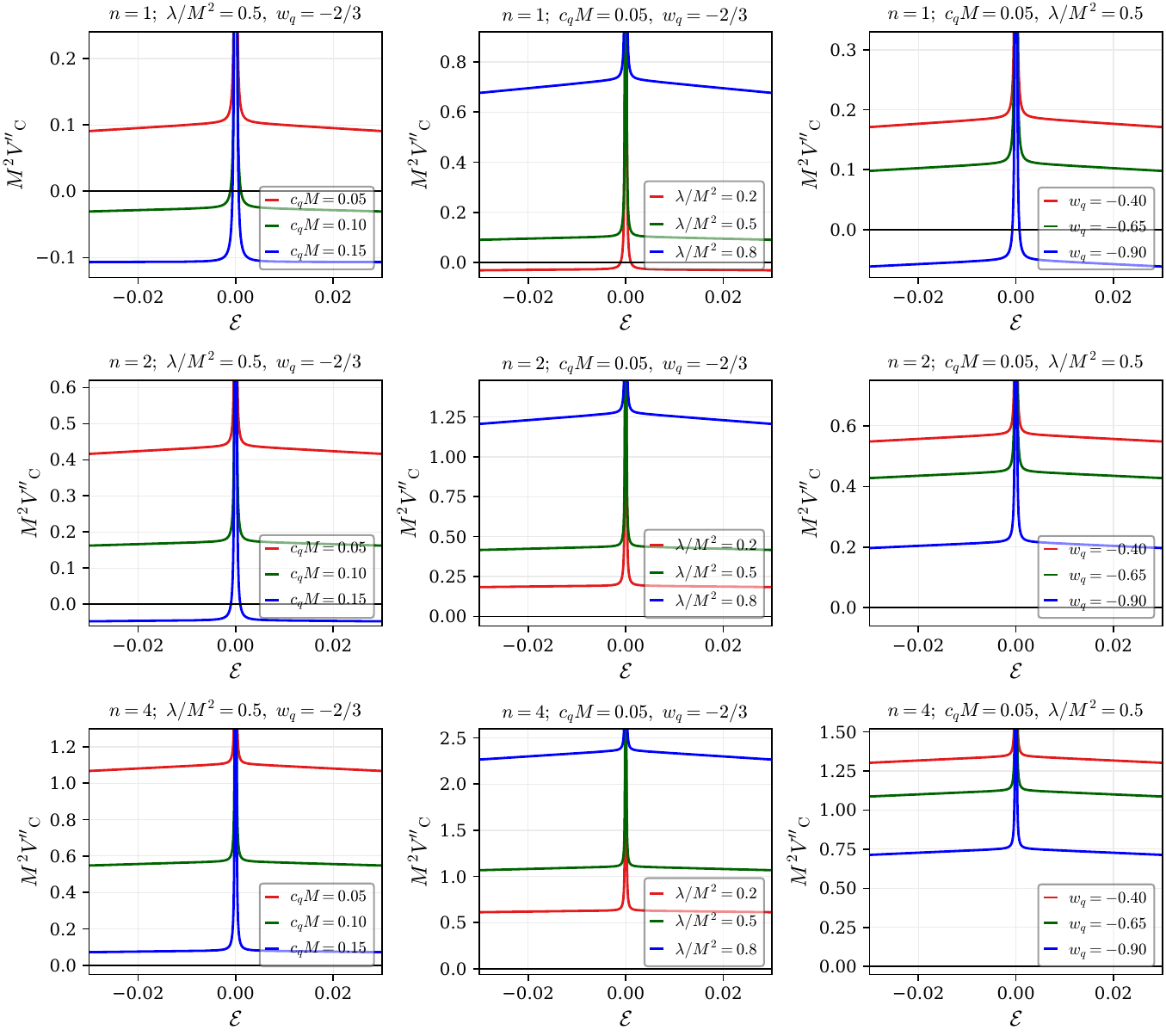}
    \caption{Variable Chaplygin stability function $M^2V''_{\mathrm{C}}(a_0)$ as a function of the dimensionless throat displacement $\mathcal{E}$, with $a_0=r_h+|\mathcal{E}|M$. The rows use $n=1,2,4$. In the first column, $\lambda/M^2=0.5$ and $w_q=-2/3$ are fixed, and the curves vary $c_qM=0.05,0.10,0.15$. In the second column, $c_qM=0.05$ and $w_q=-2/3$ are fixed, and the curves vary $\lambda/M^2=0.2,0.5,0.8$. In the third column, $c_qM=0.05$ and $\lambda/M^2=0.5$ are fixed, and the curves vary $w_q=-0.4,-0.65,-0.9$. In all panels $M=1$, with $B_C$ calibrated at each static throat before the local perturbation test. The horizontal zero level separates stable configurations, $V''>0$, from unstable ones, $V''<0$.}
    \label{fig:stability-chaplygin}
\end{figure*}

The Chaplygin panels in Fig.~\ref{fig:stability-chaplygin} display a qualitatively different stability pattern from Figs.~\ref{fig:stability-barotropic} and \ref{fig:stability-phantom}. Several branches have $M^2V''_{\mathrm C}(a_0)>0$ throughout the visible interval away from the excluded center, so the calibrated variable Chaplygin law admits local stability windows for the sampled throats. This conclusion is local in the sense of Eq.~\eqref{eq:linear-mode}: every value of $\mathcal{E}$ labels a distinct static radius and a corresponding calibrated $B_C$, not a finite-amplitude trajectory of one shell.

The row structure isolates the role of the exponent $n$. For $n=1$, the sign still depends strongly on the geometric sweep: in the first column the $c_qM=0.05$ branch is positive while larger $c_qM$ values can be negative away from the central peak; in the second column larger $\lambda/M^2$ raises the curve; and in the third column the more negative value $w_q=-0.9$ lies below the zero level outside the near-horizon spike. At $n=2$, most displayed branches are lifted into the positive region, and at $n=4$ all sampled panels remain positive on the plotted scale. The visual trend is consistent with Eq.~\eqref{eq:chaplygin-vpp}, where the explicit-radius contribution proportional to $n\sigma_0p_0$ changes the balance of the stability curvature.

The sharp central peaks should be interpreted with the same caution as in the other stability figures. They indicate the behavior of the checked curvature very close to the selected horizon, but the horizon itself is not a valid calibration point because the static density and pressure limits require separate control. Within the displayed finite windows, increasing $\lambda/M^2$ and increasing $n$ tend to favor positive curvature, whereas increasing $c_qM$ or making $w_q$ more negative can lower the branch. These trends establish sampled local stability sectors for the Chaplygin closure, not a global classification of the entire parameter space.

\section{Null geodesics and optical geometry}\label{sec:optics}

The cut-and-paste construction determines not only the surface
stress tensor and the radial response of the throat, but also the
continuation of freely propagating trajectories across the identified
hypersurface. We follow the organization used by Diemer and Smolarek
for thin-shell wormholes~\cite{DiemerSmolarek2013}, while retaining
the nonareal angular function $H(r)$ of the polymer--quintessence
geometry throughout. In this section, we first formulate the
null-geodesic problem in the two exterior copies. We then use the
corresponding effective potential and embedding geometry to classify
representative trajectories. The geodesic and embedding analyses provide the basis for the
reduced transfer calculation developed in the final part of this
section, where an infinitesimally thin equatorial disk is used to
construct face-on intensity maps.

The optical potential introduced below must not be confused with the
shell potential $V(a)$ defined in Eq.~\eqref{eq:potential}. The
latter governs radial perturbations of the material throat, whereas the
former governs test particles and photons propagating on a fixed static
wormhole background.

\subsection{Geodesic construction}
\label{subsec:geodesic-construction}

Consider a static throat at $a(\tau)=a_{0}$, located inside the connected positive-lapse region specified in
Sec.~\ref{sec:construction}. By spherical symmetry, any geodesic can be
placed in a plane through the center. Choosing this plane as the
equatorial plane, $\theta=\pi/2$, the geodesic Lagrangian may be written
as
\begin{equation}
    2\mathcal{L}_{\rm g}
    =
    -F(r)\dot{t}^{\,2}
    +
    \frac{\dot{r}^{\,2}}{F(r)}
    +
    H(r)\dot{\phi}^{\,2}
    =
    -\delta ,
    \label{eq:geodesic-lagrangian}
\end{equation}
where the overdot denotes differentiation with respect to an affine
parameter. The choices $\delta=0$ and $\delta=1$ describe null and
timelike geodesics, respectively.

Stationarity and axial symmetry give the conserved quantities
\begin{equation}
    E=F(r)\dot{t},
    \qquad
    L=H(r)\dot{\phi},
    \label{eq:geodesic-conserved}
\end{equation}
which are interpreted as the conserved Killing energy and angular
momentum per unit affine normalization. Substitution into
Eq.~\eqref{eq:geodesic-lagrangian} gives
\begin{equation}
    \dot{r}^{\,2}
    =
    E^{2}
    -
    F(r)
    \left[
        \delta+\frac{L^{2}}{H(r)}
    \right].
    \label{eq:geodesic-radial}
\end{equation}
The nonareal character of the polymer geometry appears explicitly in
the centrifugal term: $L^{2}/H(r)$ replaces the standard
$L^{2}/r^{2}$ contribution.

To describe both exterior copies by a single global coordinate, we
introduce
\begin{equation}
    r(\chi)=a_{0}+|\chi|,
    \qquad
    -\infty<\chi<\infty .
    \label{eq:global-coordinate}
\end{equation}
The throat is located at $\chi=0$, while $\chi>0$ and $\chi<0$ label the
two identical exterior regions. The areal radius
\begin{equation}
    R(\chi)
    =
    \sqrt{H\!\left(a_{0}+|\chi|\right)}
\end{equation}
reaches its minimum retained value at the shell. Reflection symmetry and the continuity of the induced metric imply that the conserved quantities associated with the tangential Killing
directions are matched continuously across the junction.

For null geodesics, it is convenient to define the impact parameter
\begin{equation}
    b\equiv \frac{L}{E}.
    \label{eq:impact-parameter}
\end{equation}
Equation~\eqref{eq:geodesic-radial} then becomes
\begin{equation}
    \dot{r}^{\,2}
    =
    E^{2}
    \left[
        1-b^{2}\frac{F(r)}{H(r)}
    \right].
    \label{eq:null-radial}
\end{equation}
The quantity
\begin{equation}
    \mathcal{P}(r)\equiv \frac{F(r)}{H(r)}
    \label{eq:photon-potential-ratio}
\end{equation}
therefore controls the radial propagation of photons.

An exterior circular null orbit satisfies
\begin{equation}
    \left.
    \frac{d}{dr}
    \left(
        \frac{F}{H}
    \right)
    \right|_{r=r_{\rm ph}}
    =0,
    \label{eq:photon-sphere-condition}
\end{equation}
or, equivalently,
\begin{equation}
    F'(r_{\rm ph})H(r_{\rm ph})
    -
    F(r_{\rm ph})H'(r_{\rm ph})
    =0.
\end{equation}
The corresponding critical impact parameter is
\begin{equation}
    b_{\rm c}^{2}
    =
    \frac{H(r_{\rm ph})}{F(r_{\rm ph})}.
    \label{eq:critical-impact}
\end{equation}
For the representative configuration used below,
\begin{equation}
    M=1,
    \qquad
    \frac{\lambda}{M^{2}}=0.5,
    \qquad
    c_{q}M=0.05,
    \qquad
    w_{q}=-\frac{2}{3},
    \qquad
    \frac{a_{0}}{M}=1.8,
    \label{eq:optical-parameters}
\end{equation}
the positive roots of $F$ are
\begin{equation}
    \frac{r_{\rm h}}{M}\simeq 1.34165,
    \qquad
    \frac{r_{\rm c}}{M}\simeq 17.87603,
\end{equation}
while
\begin{equation}
    \frac{r_{\rm ph}}{M}\simeq 2.11316,
    \qquad
    \frac{b_{\rm c}}{M}\simeq 4.72897.
    \label{eq:numerical-rph-bc}
\end{equation}
The ordering
\begin{equation}
    r_{\rm h}<a_{0}<r_{\rm ph}<r_{\rm c}
\end{equation}
places the throat inside the photon sphere while keeping it in the
positive-lapse region.

Because the wormhole and the black hole control share the same exterior
metric for $r\geq a_{0}$, they also share the same exterior photon
sphere and the same value of $b_{\rm c}$. Their optical differences do
not arise from a displacement of the critical curve, but from the
different continuation of subcritical trajectories: a black hole ray is
absorbed at the horizon, whereas a wormhole ray may cross the throat and
probe the second exterior.

\subsection{Embedding diagram}
\label{subsec:embedding}

The spatial geometry followed by an equatorial ray can be visualized on
a constant-$t$ slice. Restricting the metric to the same equatorial plane gives
\begin{equation}
    dl^{2}
    =
    \frac{dr^{2}}{F(r)}
    +
    H(r)d\phi^{2}.
    \label{eq:equatorial-spatial-metric}
\end{equation}
We embed this surface in Euclidean cylindrical coordinates
$(R,\phi,z)$, with
\begin{equation}
    dl_{\mathbb{E}^{3}}^{2}
    =
    \left[
        1+
        \left(\frac{dz}{dR}\right)^{2}
    \right]dR^{2}
    +
    R^{2}d\phi^{2}.
\end{equation}
Identifying
\begin{equation}
    R(r)=\sqrt{H(r)}
\end{equation}
and matching the radial coefficients yields
\begin{align}
    \left(\frac{dz}{dr}\right)^{2}
    &=
    \frac{1}{F(r)}
    -
    \left[
        \frac{d\sqrt{H(r)}}{dr}
    \right]^{2}
    \nonumber\\
    &=
    \frac{1}{F(r)}
    -
    \frac{r^{2}}{H(r)}.
    \label{eq:embedding-equation}
\end{align}
Choosing $z(a_{0})=0$, the two sheets are represented by
\begin{equation}
    z_{\pm}(r)
    =
    \pm
    \int_{a_{0}}^{r}
    \sqrt{
        \frac{1}{F(\bar r)}
        -
        \frac{\bar r^{2}}{H(\bar r)}
    }\,
    d\bar r .
    \label{eq:embedding-integral}
\end{equation}
The embedding is meaningful only where the square root is real and
$F>0$. The use of $R=\sqrt{H}$ is essential: plotting the surface with
$r$ as the Euclidean cylindrical radius would incorrectly identify the
seed coordinate with the physical circumference radius.

The two signs in Eq.~\eqref{eq:embedding-integral} meet at the circular
throat. The embedding diagrams shown below consequently provide a
geometric distinction between trajectories confined to one exterior
and trajectories that cross into the second copy.

\subsection{Geodesic effective potential and representative orbit classes}
\label{subsec:geodesic-potential}

In terms of the global coordinate $\chi$, the test-particle potential is
\begin{equation}
    U_{\rm geo}(\chi)
    =
    F\!\left(a_{0}+|\chi|\right)
    \left[
        \delta
        +
        \frac{L^{2}}
        {H\!\left(a_{0}+|\chi|\right)}
    \right],
    \label{eq:geodesic-potential}
\end{equation}
and the radial equation takes the form
\begin{equation}
    \dot{\chi}^{\,2}
    +
    U_{\rm geo}(\chi)
    =
    E^{2}.
    \label{eq:geodesic-energy-equation}
\end{equation}
Allowed regions satisfy $E^{2}\geq U_{\rm geo}$, while equality
identifies radial turning points, and stationary points of $U_{\rm
geo}$ can support circular orbits; the circular-orbit condition for
light, $\delta=0$, and the associated critical impact parameter were
already obtained in Eqs.~\eqref{eq:photon-sphere-condition} and
\eqref{eq:critical-impact}. For null rays, the potential reduces to
\begin{equation}
    U_{\rm null}(\chi)
    =
    L^{2}
    \frac{
        F\!\left(a_{0}+|\chi|\right)
    }{
        H\!\left(a_{0}+|\chi|\right)
    }.
    \label{eq:null-potential}
\end{equation}
The potential is even in $\chi$ because the construction is
reflection symmetric. The global coordinate $\chi$ is introduced only to represent the two
identical exterior copies in a single effective-potential diagram. The
numerical geodesic integration is performed directly in the radial
coordinate $r$, with throat-crossing rays matched continuously across
the junction. Only the ratio $L^{2}/E^{2}$ fixes the spatial
path of a null ray; separate values of $L$ and $E$ set its affine
normalization. This fact is used below to compare orbit classes
without assigning observational meaning to the exploratory energy
scale.

\begin{figure*}[htp!]
    \centering
    \includegraphics[width=0.98\linewidth]{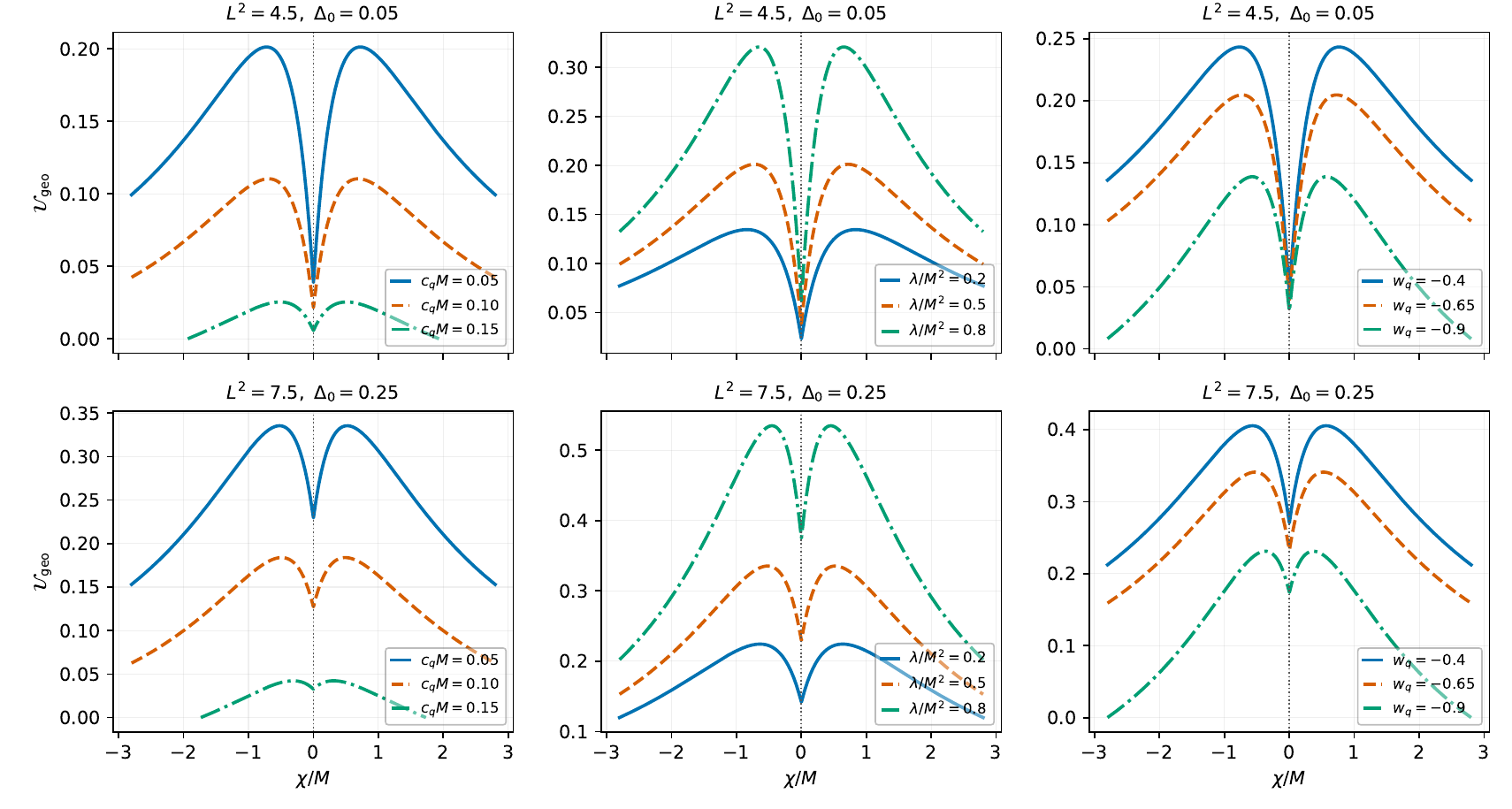}
    \caption{Null-geodesic effective potential $U_{\rm geo}$ as a
    function of the global thin-shell coordinate $\chi/M$. The upper
    row uses $L^2=4.5$ and $\Delta_0=(a_0-r_h)/M=0.05$, while the
    lower row uses $L^2=7.5$ and $\Delta_0=0.25$. The first column
    fixes $\lambda/M^2=0.5$ and $w_q=-2/3$ and varies
    $c_qM=0.05,0.10,0.15$; the second fixes $c_qM=0.05$ and
    $w_q=-2/3$ and varies $\lambda/M^2=0.2,0.5,0.8$; the third fixes
    $c_qM=0.05$ and $\lambda/M^2=0.5$ and varies
    $w_q=-0.4,-0.65,-0.9$. All panels use $M=1$ and $\delta=0$. Curves
    are restricted to their positive-lapse components, and the
    vertical dotted line marks the throat.}
    \label{fig:optical-potential}
\end{figure*}

The reflection symmetry of Fig.~\ref{fig:optical-potential} follows
directly from the dependence on $|\chi|$ in
Eq.~\eqref{eq:geodesic-potential}. Each side contains the same photon
barrier, while the central value is fixed by the position of the cut.
Moving from the upper to the lower row both increases $L^2$ and places
the throat farther from the selected smallest positive root $r_h$; the
first change raises the null barrier multiplicatively, whereas the
second raises the central potential toward the nearby barrier. These
combined row changes therefore alter the scale and the depth of the
central well without changing its two-copy symmetry.

The column comparisons isolate the geometry entering the ratio $F/H$.
Within the displayed families, increasing $c_qM$ lowers the barrier and
shortens the available static interval, increasing $\lambda/M^2$ raises
the barrier, and making $w_q$ more negative lowers it. The finite
endpoints of some curves, most visibly for the largest $c_qM$, occur
because the positive-lapse component terminates at a cosmological root;
they are not asymptotic limits. Since turning points are intersections
with a selected $E^2$, the figure identifies how the accessible orbit
classes shift across the sampled geometry but does not by itself select
a physical photon source.

We now use this potential to display representative optical
configurations. The terminology follows the standard classification
into two-world bound orbits (TWB), escape orbits that remain in the
original exterior (EO), and two-world escape orbits (TWE). These
examples are numerical diagnostics of the already defined geodesic
problem, not a separate physical model.

A two-world bound trajectory has energy below the exterior photon
barrier but high enough to cross the throat; it oscillates between
turning points located on opposite sheets. A same-side escape
trajectory is reflected at an exterior turning point and returns to the
same asymptotic side without reaching the throat. Finally, a two-world
escape trajectory has sufficient energy to cross the barrier and the
throat, subsequently propagating toward the outer region of the second
copy.

\begin{figure*}[htp!]
    \centering
    \includegraphics[width=0.98\linewidth]{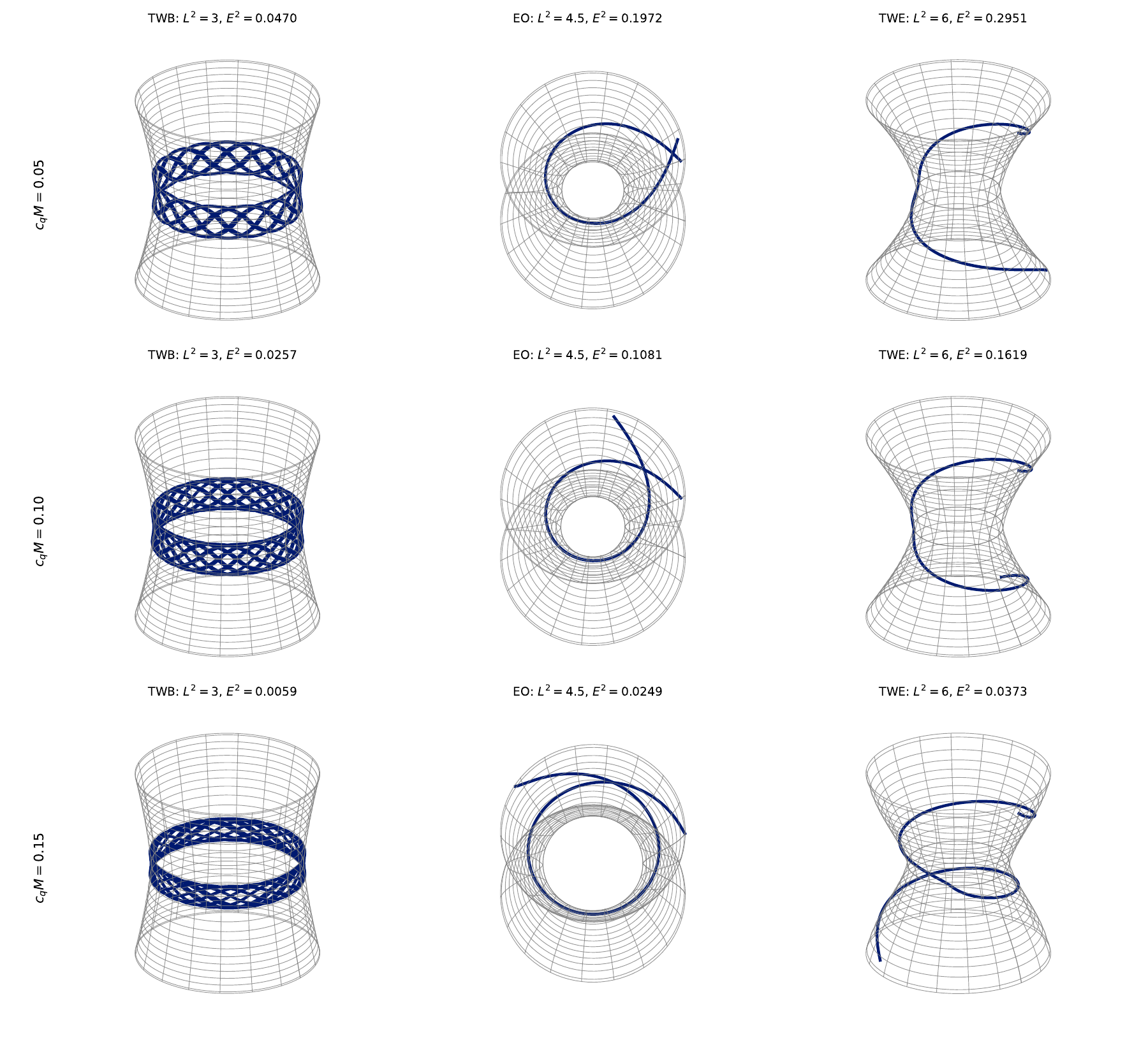}
    \caption{Equatorial Euclidean embeddings of representative null
    geodesics. The rows use $c_qM=0.05,0.10,0.15$, while $M=1$,
    $\lambda/M^2=0.5$, $w_q=-2/3$, and $\Delta_0=0.05$ are fixed. The
    columns show a two-world bound orbit (TWB) with $L^2=3$ and
    $E^2/U_{\max}=0.35$, a same-side escape orbit (EO) with $L^2=4.5$
    and $E^2/U_{\max}=0.98$, and a two-world escape orbit (TWE) with
    $L^2=6$ and $E^2/U_{\max}=1.10$. Here $U_{\max}$ is the maximum
    null barrier in the corresponding panel, and the resulting value
    of $E^2$ is printed above each diagram. The grey wireframe is the
    embedded slice and the dark curve is the single geodesic displayed
    in each panel.}
    \label{fig:embedded-geodesics}
\end{figure*}

Figure~\ref{fig:embedded-geodesics} translates the turning-point
structure of Eq.~\eqref{eq:geodesic-energy-equation} into the two-sheet
geometry. In the first column the energy lies below the photon barrier
but above the throat value, so the ray repeatedly winds while remaining
bounded between turning points on opposite sheets. The middle-column
ray is reflected by the outer barrier and remains in its original
exterior. In the last column the energy exceeds the barrier, allowing
the trajectory to pass through the throat and escape into the second
exterior. The panels therefore distinguish TWB, EO, and TWE behavior by
topology and turning points, rather than by the visual orientation of
the three-dimensional view.

Increasing $c_qM$ down the rows reduces the barrier scale, which
explains why the printed $E^2$ values decrease when the fixed ratios
$E^2/U_{\max}$ are maintained. The separately framed surfaces should
not be used for an absolute size comparison, but they retain the
correct areal radius and embedding slope from
Eq.~\eqref{eq:embedding-equation}. For the sampled configurations, all
three orbit classes persist as the Kiselev amplitude changes. This is a
finite numerical classification of representative null rays, not a
claim that every admissible throat supports the same complete orbit
inventory.

\subsection{Face-on thin equatorial disk}
\label{subsec:thin-disk}

The embedding diagrams provide an intuitive geometric representation
of selected trajectories. To construct a corresponding image, we now label the null geodesics by their impact parameter on the observer's screen and determine their intersections
with an emitting surface.

Although the generic geodesic equations above were written after rotating each orbit to an equatorial orbital plane, a face-on ray from the symmetry axis to the disk lies in a meridional plane. By spherical symmetry, this change of orbital-plane orientation leaves the radial dynamics and the in-plane angular evolution unchanged. The face-on transfer calculation may therefore use the same geodesic relations in the meridional plane.

We consider an infinitesimally thin equatorial disk and place the observer on its symmetry axis. The observer therefore sees
the disk face-on. Since the spacetime is spherically symmetric and the
emitting configuration is axially symmetric around the line of sight,
the observed intensity depends only on the radial screen coordinate
\begin{equation}
    b=\sqrt{\alpha^{2}+\beta^{2}}.
    \label{eq:screen-radius}
\end{equation}
Here $(\alpha,\beta)$ denote impact-parameter coordinates rather than locally measured angular coordinates on the observer's screen. Since the Kiselev-type geometry is not asymptotically flat, the conserved impact parameter defined in Eq.~\eqref{eq:impact-parameter} is used as the radial image coordinate throughout this work. Thus, the observer is placed at a finite radius $r_{\rm obs}$ inside the static patch. In the numerical implementation we take
\begin{equation}
    r_{\rm obs}
    =
    r_{\rm h}
    +
    0.92\left(r_{\rm c}-r_{\rm h}\right),
\end{equation}
which gives $r_{\rm obs}/M\simeq 16.5533$ for
Eq.~\eqref{eq:optical-parameters}.

For a static observer located at the finite radius
$r_{\rm obs}$, this impact parameter provides a convenient label for null
geodesics but should not be identified directly
with a locally measured angular coordinate.
Introducing the observer's orthonormal tetrad,
the local viewing angle $\Theta$ satisfies
\begin{equation}
\sin\Theta
=
b
\sqrt{\frac{F(r_{\rm obs})}{H(r_{\rm obs})}}.
\label{eq:local-angle}
\end{equation}
Accordingly, the images presented below are
constructed in impact-parameter coordinates.
Since all three configurations use the same
observer position and the same exterior
geometry, the monotonic mapping between
$b$ and the local viewing angle does not
modify the relative ring morphology. The maximum impact parameter sampled
at this position is restricted by
\begin{equation}
    b^{2}
    <
    \frac{H(r_{\rm obs})}{F(r_{\rm obs})}.
\end{equation}

We model the radial surface emission of the equatorial disk using the profile introduced by Gralla, Lupsasca, and Marrone \cite{Gralla2020PhotonRing} and subsequently adopted for the polymer--quintessence source geometry \cite{Araujo2025}:
\begin{equation}
    I_{\rm em}(r;\mu)
    =
    A\,
    \frac{
        \exp\left\{
        -\frac{1}{2}
        \left[
            \gamma+
            \operatorname{arcsinh}
            \left(
                \frac{r-\mu}{\varsigma}
            \right)
        \right]^{2}
        \right\}
    }{
        \sqrt{(r-\mu)^{2}+\varsigma^{2}}
    },
    \label{eq:disk-emissivity}
\end{equation}
Here $I_{\rm em}$ denotes the emitted bolometric surface intensity, $A$ is a normalization constant, $\mu$ determines the radial
location of the emitting profile, $\varsigma$ controls its radial width,
and $\gamma$ controls its asymmetry. In the images below,
\begin{equation}
    \frac{\varsigma}{M}=\frac{1}{8},
    \qquad
    \gamma=-2.
\end{equation}
Two physically distinct choices are considered for the wormhole:
$\mu=r_{\rm ph}$, which concentrates the emissivity around the exterior
photon-sphere region, and $\mu=a_{0}$, which concentrates it near the
throat. The black hole control uses $\mu=r_{\rm ph}$.

For a null ray, Eqs.~\eqref{eq:geodesic-conserved} and
\eqref{eq:null-radial} give the absolute angular advance
\begin{equation}
    \left|
        \frac{d\psi}{dr}
    \right|
    =
    \frac{
        b
    }{
        H(r)
        \sqrt{
            1-b^{2}F(r)/H(r)
        }
    }.
    \label{eq:angular-propagation}
\end{equation}
The angle $\psi$ is measured in the plane containing the ray and the
symmetry axis, with $\psi=0$ at the observer. The ray crosses the
equatorial disk whenever
\begin{equation}
    \psi_{k}
    =
    \frac{\pi}{2}+k\pi,
    \qquad
    k=0,1,2,\ldots .
    \label{eq:disk-crossings}
\end{equation}
Here $\psi_k$ is the angular advance at the $k$th disk crossing, and $k$ is a nonnegative integer that orders successive crossings along the ray.

For each impact parameter $b$, the radii $r_{k}(b)$ satisfying
Eq.~\eqref{eq:disk-crossings} are obtained by integrating
Eq.~\eqref{eq:angular-propagation} along all available radial branches.

The propagation differs qualitatively on the two sides of the critical
curve. For the black hole control, a subcritical ray with $b<b_{\rm c}$
follows a plunging trajectory that terminates at the event
horizon, whereas a ray with $b>b_{\rm c}$ reaches an exterior turning
point and returns toward the observer-side outer region. For the
wormhole, a subcritical ray is not absorbed. It reaches the throat and
continues into the second exterior. Supercritical rays turn around
outside the throat and remain in the observer-side exterior.

The limiting case $b=b_{\rm c}$ asymptotically approaches the unstable
circular photon orbit. Numerically, the impact-parameter grid is adaptively refined in the vicinity of $b_{\rm c}$ to accurately resolve the logarithmic divergence of the orbital angle associated with the unstable photon orbit.

\subsection{Transfer prescription and emitted-power matching}
\label{subsec:transfer}

The observed contribution of each disk crossing follows from the
invariance of $I_{\nu}/\nu^{3}$ along a null geodesic. For a static
emitter and a static observer, the frequency-shift factor is
\begin{equation}
    g_{k}
    =
    \frac{\nu_{\rm obs}}{\nu_{\rm em}}
    =
    \sqrt{
        \frac{
            F(r_{k})
        }{
            F(r_{\rm obs})
        }
    }.
    \label{eq:redshift-factor}
\end{equation}
For frequency-integrated, bolometric surface intensity, the
transformation is proportional to $g^{4}$. Thus
\begin{equation}
    I_{\rm obs}(b)
    =
    \sum_{k}
    g_{k}^{4}
    I_{\rm em}\!\left(r_{k}(b)\right).
    \label{eq:observed-intensity-general}
\end{equation}
Since $F(r_{\rm obs})$ is common to all crossings and all three models,
the numerical profiles may equivalently be written, up to a global
constant, as
\begin{equation}
    I_{\rm obs}(b)
    \propto
    \sum_{k}
    F\!\left(r_{k}(b)\right)^{2}
    I_{\rm em}\!\left(r_{k}(b)\right).
    \label{eq:observed-intensity}
\end{equation}
This is a discrete transfer sum, not a volumetric line-of-sight
integral. Multiple images arise whenever one geodesic crosses the disk
more than once.

A direct comparison between the one-sided black hole model and the
two-sided wormhole requires an emissivity normalization. Otherwise, the
wormhole would be brighter simply because it contains two identical
emitting exteriors. To isolate the geometrical and topological contributions from this
trivial source-counting effect, we match the total emitted power.

On a constant-$t$ equatorial slice, the proper area element of one disk
is
\begin{equation}
    dA_{\rm prop}
    =
    2\pi
    \sqrt{
        \frac{H(r)}{F(r)}
    }
    dr.
    \label{eq:disk-proper-area}
\end{equation}
The total locally emitted disk power is consequently taken as
\begin{equation}
    \mathscr{P}_{\rm loc}
    \propto
    N_{\rm s}
    \int_{r_{\rm min}}^{r_{\rm obs}}
    I_{\rm em}(r)
    \,
    2\pi
    \sqrt{
        \frac{H(r)}{F(r)}
    }
    dr,
    \label{eq:local-disk-power}
\end{equation}
Here $\mathscr{P}_{\rm loc}$ denotes the total locally emitted power summed over the emitting exterior copies, while $N_{\rm s}=1$ for the black hole control and $N_{\rm s}=2$ for
the symmetric wormhole. The lower limit is $r_{\rm h}$ in the
black hole case and $a_{0}$ for each wormhole exterior.

We also consider a redshift-weighted emitted power,
\begin{equation}
    \mathscr{P}_{\rm K}
    \propto
    N_{\rm s}
    \int_{r_{\rm min}}^{r_{\rm obs}}
    F(r)I_{\rm em}(r)
    \,
    2\pi
    \sqrt{
        \frac{H(r)}{F(r)}
    }
    dr.
    \label{eq:killing-disk-power}
\end{equation}
The notation $\mathscr{P}_{\rm K}$ emphasizes that this quantity is associated
with the static Killing-time redshift. Because the spacetime has a
finite outer static boundary and is not asymptotically flat, it should
not literally be interpreted as luminosity measured at spatial
infinity.

The figure discussed below uses the equal-$\mathscr{P}_{\rm loc}$
normalization. Accordingly, the amplitudes of both wormhole disks are
reduced so that the sum of the power emitted in the two copies equals
the power of the single black hole disk.

\subsection{Image morphology and wormhole--black hole comparison}
\label{subsec:image-morphology}

\begin{figure*}[htp!]
    \centering
    \includegraphics[width=0.98\linewidth]{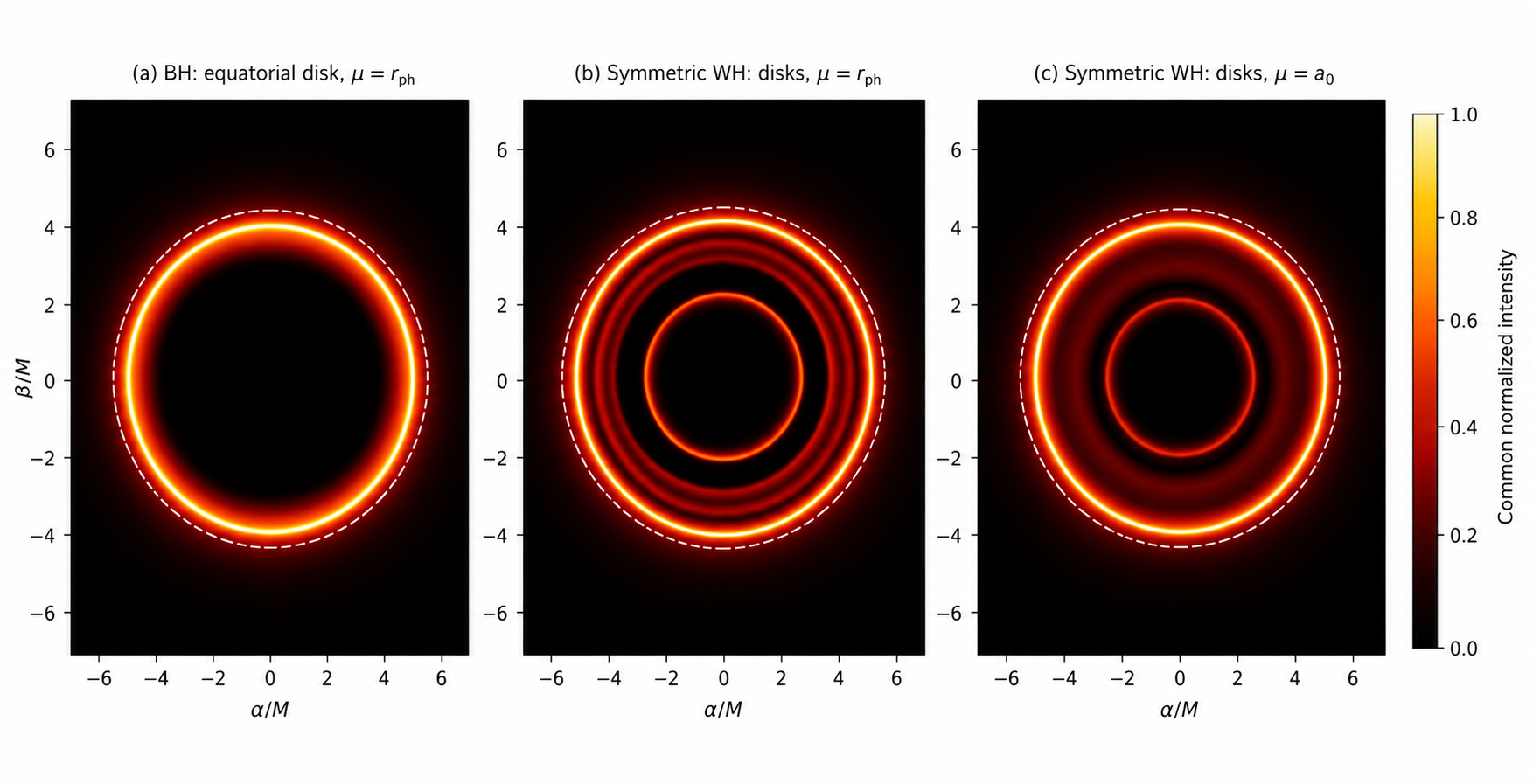}
    \caption{ Face-on images of an infinitesimally thin equatorial emitting disk
    for the representative polymer--quintessence configuration
    $M=1$, $\lambda/M^{2}=0.5$, $c_{q}M=0.05$,
    $w_{q}=-2/3$, $a_{0}/M=1.8$, $\varsigma/M=0.125$, and
    $\gamma=-2$. The observer lies on the disk symmetry axis.
    Panel (a) shows the black hole control with one emitting exterior
    and $\mu=r_{\rm ph}$. Panel (b) shows the symmetric thin-shell
    wormhole with identical disks in the two exterior copies and
    $\mu=r_{\rm ph}$. Panel (c) shows the same wormhole with the
    emissivity concentrated closer to the throat, $\mu=a_{0}$.
    The disk amplitudes are fixed by equal total locally emitted power,
    Eq.~\eqref{eq:local-disk-power}. The dashed circle marks the common
    critical curve $b=b_{\rm c}$, with
    $b_{\rm c}/M\simeq4.72897$. The color scale represents a common
    normalization of the three intensity profiles. The displayed maps
    are post-processed visualizations of the computed radial
    geodesic-transfer profiles; quantitative statements concerning
    ring positions are obtained from $I_{\rm obs}(b)$ rather than from
    the displayed ring thickness.}
    \label{fig:thin-disk-images}
\end{figure*}

Figure~\ref{fig:thin-disk-images} compares the black hole imaging control with two symmetric wormhole configurations. The dashed circle marks the common critical impact parameter $b=b_{\rm c}$. Since all three models share the same exterior metric and exterior photon sphere, the dashed circle occupies the same radial position in every panel. The images are intended as qualitative synthetic comparisons between the black hole and wormhole spacetimes using the same thin-shell emission model.

Panel~\ref{fig:thin-disk-images}(a) is dominated by a broad direct
emission feature and a narrow enhancement close to the critical curve, according to \cite{Araujo2025}.
For subcritical impact parameters, rays may intersect the disk before
being absorbed at the black hole horizon, but there is no second
exterior capable of supplying additional crossings. The black hole image is therefore dominated by a broad annular
component, together with a narrow enhancement near the critical curve.

The wormhole images are qualitatively richer. In
Fig.~\ref{fig:thin-disk-images}(b), the disk emissivity is centered at
$r_{\rm ph}$ in both copies. Rays close to $b_{\rm c}$ accumulate a
large orbital angle and can satisfy
Eq.~\eqref{eq:disk-crossings} several times. Each crossing contributes
a term to Eq.~\eqref{eq:observed-intensity}. This produces a hierarchy
of direct and higher-order annular features.

For subcritical wormhole rays, an additional difference becomes decisive. Instead
of terminating at a horizon, the backward-traced wormhole ray crosses
the throat and continues into the second exterior, where it may
intersect the second emitting disk. These second-exterior
contributions populate impact parameters that are dark or much weaker
in the black hole control. The resulting inner rings are therefore a
direct consequence of the causal continuation through the throat and
of the presence of an emitting region in the second copy.

A closely related cross-throat mechanism was recently demonstrated by Macedo \textit{et al.} for reflection-asymmetric thin-shell wormholes illuminated by one or two thin disks, where propagation into the opposite exterior produces a multiphoton-ring structure \cite{Macedo2026Multiphoton}. The present construction differs in being reflection symmetric, in using a polymer--quintessence exterior, and in matching the total emitted power of the one- and two-sided configurations. Accordingly, the additional rings found here should be viewed as a model-specific realization of this broader cross-throat imaging mechanism rather than as a generic optical signature established for every thin-shell wormhole.

The multiple-ring structure should not be interpreted as several
independent photon spheres. The representative geometry has one
exterior unstable circular photon orbit on each identical copy, and
both have the same radius. The distinct rings instead correspond to
different disk-crossing orders and different geodesic branches. Their
radial accumulation near $b_{\rm c}$ reflects the increasingly large
deflection of rays approaching the unstable photon orbit.

Panel~\ref{fig:thin-disk-images}(c) retains the same spacetime and geodesic structure but shifts the peak of the disk emissivity from $\mu=r_{\rm ph}$ to $\mu=a_{0}$. The critical curve therefore remains unchanged, as it is determined solely by the exterior geometry, whereas the distribution of brightness is significantly modified. When the emissivity is concentrated near the photon sphere, higher-order trajectories—which linger near the unstable circular orbit and undergo multiple disk crossings—are efficiently illuminated, producing a richer hierarchy of subrings. Shifting the emissivity toward the throat reduces the weight of these near-critical trajectories, suppressing part of this fine structure while preserving the underlying lensing geometry. This clearly separates two independent effects: the optical geometry, which determines the allowed image locations, and the source emissivity, which determines how strongly those locations are illuminated.

The bright maximum need not coincide exactly with the dashed critical
curve. The latter is a purely geodesic quantity fixed by
Eq.~\eqref{eq:critical-impact}, whereas the peak of $I_{\rm obs}(b)$
also depends on the emissivity profile, the crossing radii, the
redshift factor, and the number of disk intersections. For the adopted
profiles, however, the dominant outer peak remains numerically very
close to $b_{\rm c}$ in all three cases.

Although the black hole and wormhole models share the same exterior
critical curve, they differ markedly in their image morphology. The
wormhole admits cross-throat propagation and emission from the second
exterior, generating additional image branches that are absent or
strongly suppressed in the black hole case. The present comparison is
performed within a simplified transfer model consisting of a static,
infinitesimally thin, face-on disk, neglecting Doppler boosting,
aberration, absorption, scattering, magnetic fields, and radiative
backreaction.

Within the present transfer prescription, the second emitting exterior
can generate additional crossing branches and associated inner-ring
features. Their visibility, brightness, and contrast depend on the
emissivity profile, the relative normalization of the two disks, and
the adopted ray-transfer prescription. The present images therefore
provide a model-dependent qualitative illustration of the optical
effects of the underlying spacetime geometry.

\section{Conclusion}\label{sec:conclusion}

In this work, we constructed a reflection-symmetric thin-shell wormhole by gluing two copies of a connected positive-lapse sector of the polymer-quintessence geometry and asked how the nonareal angular structure of the seed metric reshapes the standard thin-shell analysis. The construction shows that the throat is controlled by the true area function rather than by the seed radial coordinate itself, and this distinction propagates through the junction conditions, the conservation law, the radial dynamics, and the optical sector. In particular, the shell balance acquires a momentum-flux contribution that survives in the pure-polymer limit and disappears only in the areal-radius limit, so the hybrid geometry cannot be reduced consistently to the usual transparent-shell bookkeeping by using reciprocal temporal and radial coefficients alone.

At the level of the surface matter, every static throat on the retained positive radial branch carries negative surface energy density, while the sign of the tangential null-energy and intrinsic strong-energy combinations remains controlled by the local slope of the lapse. The local radial dynamics can nevertheless be organized in a general effective-potential form that keeps both the nonareal flux correction and the explicit radius dependence of the surface equation of state. Within the sampled configurations examined here, the calibrated linear barotropic and variable phantomlike closures do not generate positive local curvature of the shell potential, whereas the variable Chaplygin law does open finite local stability sectors, especially when the explicit radial contribution becomes sufficiently strong. These statements remain local to the calibrated static configurations and to the explored parameter windows; they do not establish global stability of the full families.

The thermodynamic and optical analyses sharpen the physical interpretation of this geometry beyond the purely mechanical test. The shell admits a well-defined local acceleration temperature, but its fixed-parameter first-law balance is modified by the same polymer contribution that appears in the conservation equation, so the entropy bookkeeping is more restrictive than in areal-radius constructions. In the optical sector, the null-geodesic potential and the associated embedding diagrams are complemented by simplified thin-disk images showing that, although the wormhole and the corresponding black hole geometry share the same exterior critical curve, cross-throat propagation generates additional inner image branches in this transfer setup.

The scope of these conclusions should be kept within the assumptions of the present model. Our stability test is linear, radial, and local around static symmetric throats; the constitutive laws are phenomenological surface closures; the numerical evidence comes from finite parameter scans; and neither shell microphysics nor nonlinear backreaction has been incorporated. For the same reason, the thermodynamic discussion is a local bookkeeping analysis rather than a complete equilibrium phase structure, and the optical discussion stops short of full ray tracing or observable image construction.

Natural extensions are therefore clear. A broader parameter survey could map the local stability sectors more systematically, especially for the Chaplygin branch; asymmetric junctions and nonradial or nonlinear perturbations could test how much of the present picture survives beyond the reflection-symmetric setting; and more realistic radiative-transfer calculations, including accretion dynamics, Doppler boosting, absorption, and magnetized plasma effects, could connect the simplified optical images presented here to astrophysical observations. Taken together, the present results show that the combination of polymer corrections and a quintessence environment does more than deform a known seed metric: it reorganizes the shell mechanics, thermodynamic bookkeeping, and optical appearance of the thin-shell wormhole built from it.

\begin{acknowledgments}
EO thanks the Fundação Cearense de Apoio ao Desenvolvimento Científico e Tecnológico (FUNCAP), through grant BP6-0241-00335.01.00/25. MCA would like to thank FUNCAP, Funda\c{c}\~{a}o Cearense de Apoio ao Desenvolvimento Cient\'{i}fico e Tecnol\'{o}gico (Process No. DC3-0235-00076.01.00/24), and CNPq, Conselho Nacional de Desenvolvimento Cient\'{i}fico  e Tecnol\'{o}gico - Brasil (Process No. 304145/2025-4), for financial support. JF would like to thank Funda\c{c}\~{a}o Cearense de Apoio ao Desenvolvimento Cient\'{i}fico e Tecnol\'{o}gico (FUNCAP) under the grant PRONEM PNE0112- 00085.01.00/16 and the Conselho Nacional de Desenvolvimento Científico e Tecnol\'{o}gico (CNPq) under the grant 304485/2023-3. CRM would like to thank the Conselho Nacional de Desenvolvimento Cient\'{i}fico e Tecnol\'{o}gico (CNPq) for partial financial support, through grant 301122/2025-3.
\end{acknowledgments}

\bibliography{ref}

\end{document}